\documentclass[12pt]{article}

\usepackage{newtxtext,newtxmath}

\usepackage{graphicx}
\usepackage{xcolor}

\usepackage[letterpaper,margin=1in]{geometry}

\renewenvironment{abstract}
	{\quotation}
	{\endquotation}

\date{}

\makeatletter
\renewcommand{\fnum@figure}{\textbf{Figure \thefigure}}
\renewcommand{\fnum@table}{\textbf{Table \thetable}}
\makeatother

\usepackage{scicite}
\usepackage{caption}
\usepackage{url}

\newcommand{\dom}{\Delta\omega}

\newcommand{\dOm}{\Delta\Omega}
\newcommand{\twocol}[1]{#1}

\newcommand{\cb}[1]{#1}

\newcommand{\maxdim}{N}

\newcommand{\vth}{\vartheta}

\newcommand{\e}{\varepsilon}
\newcommand{\eps}{\e}

\newcommand{\abs}[1]{\left|#1\right|}

\newcommand{\sset}[1]{\left\lbrace #1\right\rbrace}

\def\scititle{
	Frequency Synchronization Induced by Frequency Detuning
}
\title{\bfseries \boldmath \scititle}

\author{
	Jorge Luis Ocampo-Espindola$^{1,2,3}$,
	Christian Bick$^{4,5,6,7}$,
	Adilson E. Motter$^{2,3,8,9}$\and
	Istv\'{a}n Z. Kiss$^{1\ast}$\and
	\small$^{1}$Department of Chemistry, Saint Louis University,  St.~Louis, MO 63103, USA.\and
	\small$^{2}$Center for Network Dynamics, Northwestern University, Evanston, IL 60208, USA.\and
	\small$^{3}$Department of Physics and Astronomy, Northwestern University, Evanston, IL 60208, USA.\and
	\small$^{4}$Department of Mathematics, Vrije Universiteit Amsterdam, Amsterdam, the Netherlands.\and
	\small$^{5}$Mathematical Institute, University of Oxford, Oxford, UK.\and
	\small$^{6}$Department of Mathematics, University of Exeter, Exeter, UK.\and
	\small$^{7}$Institute for Advanced Study, Technische Universit\"at M\"unchen, Garching, Germany.\and
	\small$^{8}$Department of Engineering Sciences and Applied Mathematics, Northwestern University, Evanston, IL 60208, USA.\and
	\small$^{9}$Northwestern Institute on Complex Systems, Northwestern University, Evanston, IL 60208, USA.\and
	\small$^\ast$Corresponding author. Email: istvan.kiss@slu.edu.
}

\begin{document} 

\maketitle

\begin{abstract} \bfseries \boldmath
Abstract: It is widely held that identical systems tend to behave similarly under comparable conditions. Yet, for systems that interact through a network, symmetry breaking can lead to scenarios in which this expectation does not hold. Prominent examples are chimera states \cb{in multistable phase-oscillator networks.} Here, we show that for a broad class of such networks,  asynchronous states can be converted into frequency-synchronized states when identical oscillators are detuned to have different intrinsic frequencies. Importantly, we show that frequency synchronization is achieved over a range of intrinsic frequency detuning and is thus a robust effect. These results, which are supported by theory, simulations, and electrochemical oscillator experiments, reveal a counterintuitive opportunity to use parameter heterogeneity to promote synchronization.

\end{abstract}

\subsection*{Introduction}
Synchronization of coupled oscillators is an emergent network phenomenon, which enjoys remarkable properties and has far-reaching implications in natural and engineered systems~\cite{Pikovsky2003,Strogatz2004}. Among the factors governing this phenomenon, coupling takes center stage as it can stabilize otherwise unstable synchronization states~\cite{arenas2008synchronization}. But coupling can also lead to instabilities and impede synchronization even when the oscillators are identical and identically coupled~\cite{Panaggio2015,omel2018mathematics}. A fundamental problem of current interest is the interplay between the coupling network and other factors governing synchronization, which include the initial conditions and oscillator parameters~\cite{wiley2006size,Martens2015,Sebek:2019gy}. The former plays a central role in the emergence of a rich variety of spatiotemporal patterns \cite{wiley2006size,abrams2016introduction}, including many instances of cluster synchronization and chimera states \cite{nicolaou2019multifaceted,matheny2019exotic}. An outstanding question in this context concerns the impact of oscillator parameters on such patterns. 

Intuitively, parameter heterogeneity across oscillators may be expected to inhibit synchronization.
This expectation is supported by studies of simple phase oscillator models, where
 frequency detuning  is observed to increase the coupling threshold for synchronization \cite{Kuramoto} 
 or even destroy synchronization \cite{Ashwin2006}. 
However, recent research on higher-dimensional oscillator models shows evidence
that heterogeneity can in fact serve as a stabilizing
element in synchronization  \cite{nishikawa2016symmetric,zhang2021random}.
This is different from earlier work on synchronization enhancement, which was centered on the manipulation of coupling network \cite{barahona2002synchronization}, control of dynamical variables \cite{yu2009pinning}, and homogenization of oscillator inputs \cite{motter2005network}.
There are now  various 
studies showing the beneficial impact of heterogeneity in applications as diverse as parametric wave amplification \cite{perego2022synchronization}, persistency of oscillations in active matter \cite{yang2022emergent}, chaos control in multilayer networks \cite{medeiros2021asymmetry}, cluster synchronization \cite{bolotov2024breathing},
state stabilization in optical resonators \cite{garbin2020asymmetric},  phase-locking in laser arrays \cite{nair2021using}, neural computation \cite{gast2024neural}, and even quantum dynamics \cite{lorch2017quantum}.
It remains an open question, however, whether frequency detuning can steer spatiotemporal patterns and promote global synchronization 
in phase oscillator networks. A phase oscillator description would offer promising new directions for theoretical advancements\cite{Acebron2005}, enabling extensions to novel types of networks and oscillators. It would also open avenues for experiments where phase models have proven effective \cite{Wiesenfeld1998,doi:10.1126/sciadv.abb2637,matheny2019exotic} in exploring emerging spatiotemporal pattern formations.

In this paper, \cb{starting from states} 
 in which identically coupled identical oscillators do not synchronize, we show how frequency synchronization can be achieved robustly when the phase oscillators are detuned. Our results are established for \cb{multistable networks of} weakly coupled oscillators that, upon phase reduction, exhibit second harmonic coupling and phase frustration. 
 \cb{Our approach leads to the synchronization of otherwise asynchronous states that coexist with synchronous ones.}
The results are further demonstrated for neuronal and chemical oscillators, with experimental validation conducted specifically for electrochemical systems. For concreteness, \cb{we first focus} on modular networks consisting of two populations of $N$ oscillators each, with all-to-all intra-population coupling of strength~$K>0$ and weaker all-to-all inter-population coupling of strength $\eps K\ge0$ (as considered in numerous previous studies \cite{abrams2008solvable,tinsley2012chimera,martens2013chimera,panaggio2016chimera,bick2017robust}). 
Notably, we show that the effect can be achieved by manipulating a single independent parameter in each population, making it scalable to large networks. This represents \cb{an important} step toward reducing computational complexity, which could otherwise require specific parameter adjustments of individual oscillators.

\begin{figure} 
\centering
\includegraphics[width= 0.9\columnwidth]{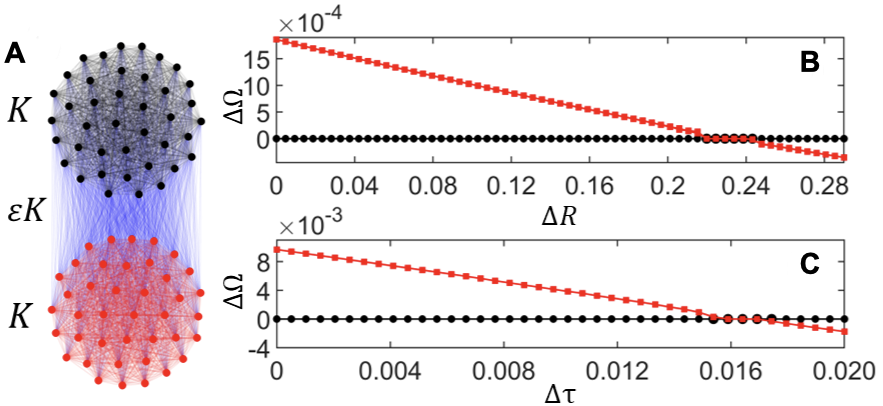} 
\caption{\textbf{ Detuning-induced frequency synchronization for two oscillator network models. }
(\textbf{A}) Modular network exhibiting all-to-all  intra- and inter-population coupling of strengths $K$ and $\eps K$, respectively.
(\textbf{B} and \textbf{C})  Frequency of individual oscillators in population 1 (black) and population 2 (red), relative to oscillator 1 in population 1, as a function of the detuning parameter for: chemical oscillators with electrical coupling in (\textbf{B}) and the integrate-and-fire model with a refractory period in (\textbf{C}). 
The detuning is introduced in  resistances ($\Delta R$) in (\textbf{B}) and relaxation timescales ($\Delta \tau$) in (\textbf{C}),
in both cases for  $N=2$ oscillators in each population. For model description and other parameters, see Supplementary Text.} 
\label{fig:ni_int}

\end{figure} 

The essence of the phenomenon is illustrated in Fig.~\ref{fig:ni_int} for two oscillator models, namely a kinetic chemical oscillator model~\cite{Kiss:2005PM} 
and an integrate-and-fire neuronal model~\cite{Kori2018}, for $N=2$.  In both cases, when the oscillators are identical, the system exhibits a type of cluster-synchronized state---also known as a weak chimera~\cite{Ashwin2014a,Bick2015c}---in which each population is synchronized to a different frequency. However, the frequency difference is gradually reduced as a parameter effectively controlling the intrinsic frequencies is varied in one of the populations.  Global frequency synchronization is then achieved when the intrinsic frequencies are sufficiently detuned between the two populations.  Importantly, this state, in which the two populations exhibit identical frequencies, is stable not just for a specific parameter assignment but rather over a range of intrinsic frequency detuning. Motivated by such simulation results, here we first establish a theory to identify the mechanisms and conditions that can give rise to this effect in modular networks of arbitrary size. \cb{We also demonstrate the robustness of the effect in modular networks with more general structures, including variations in coupling density, population number, and population sizes.} We then report on experiments in electrochemical oscillator networks to test and expand on our predictions in a realistic setting.

\subsection*{Results}
\subsubsection*{Phase-model theory predicts detuning-induced synchronization}
The theory is established using a phase oscillator model, which is a suitable mathematical description for the dynamics of 
a broad class of weakly coupled systems \cite{Kuramoto,strogatz2000kuramoto},  
including those in our simulations and experiments~\cite{Kori2018,Kiss:2005PM}. 
To elucidate the dynamical mechanisms that give rise to the emergence of frequency synchronization, we consider two-population networks 
with $\maxdim=2n$ oscillators in each population for arbitrary $n\ge 1$  (Fig.~\ref{fig:ni_int}A).
The state of oscillator $k\in\sset{1, \dotsc,\maxdim}$ in population $\sigma\in\sset{1,2}$---or simply oscillator~$(\sigma,k)$---is given by a phase variable~$\theta_{\sigma}^k$.
The oscillator phases evolve according to 
\begin{equation} 
\dot\theta_{\sigma}^k = \omega_{\sigma} + \frac{K}{\maxdim}\sum_{j=1}^N g(\theta_{\sigma}^j-\theta_{\sigma}^k)
+\frac{\eps K}{\maxdim}\sum_{j=1}^N g(\theta_{\kappa}^j-\theta_{\sigma}^k),
\label{eq:OscPopulations}
\end{equation} 
where
$\kappa = 1+(\sigma\mbox{ mod } 2)$,
$\omega_\sigma$ is the intrinsic frequency of the oscillators in population~$\sigma$, and $g$ is a $2\pi$-periodic function mediating the interaction between the oscillators. 
We focus on $0\leq \eps < 1$ and assume, without loss of generality, that
 $K=2$.
 
Consider the trajectory $\theta_{\sigma}^k (t)$ 
for an initial condition $\big(\theta_1^1(0),\dots,\theta_1^{N}(0),\theta_2^1(0),\dots, \theta_2^{N}(0)\big)=:\theta_{(0)}$.
The asymptotic average angular frequency along the trajectory is $\Omega_{\sigma}^k (\theta_{(0)})= \lim_{T\to\infty}\frac{1}{T}\theta_{\sigma}^k (T)$. 
Oscillators $(\sigma,k)$ and $(\kappa,j)$ are frequency synchronized for the trajectory with initial condition~$\theta_{(0)}$ if $\Omega_{\sigma}^k(\theta_{(0)})=\Omega_{\kappa}^j(\theta_{(0)})$.

Since the oscillators in each population are identical and also identically coupled, there are symmetries that restrict the dynamics of Eq.~\eqref{eq:OscPopulations}~\cite{Ashwin1992,Golubitsky2002}.
These symmetries imply that all oscillators within a single population have the same average angular frequency
$\Omega_{\sigma}^k(\theta_{(0)})=:\Omega_{\sigma}(\theta_{(0)})$ $\forall k$ \cite{Ashwin2014a,Bick2015d}. 
This allows us to define the frequency difference between populations as
$\Delta\Omega_{21}(\theta_{(0)}) := \Omega_{2}(\theta_{(0)})-\Omega_{1}(\theta_{(0)})$. 
The symmetries of the system also give rise to invariant subsets in which the oscillators in each population are partitioned into clusters of identical phases.
In particular,  a state in which population~$1$ is fully phase synchronized (all oscillators have phase~$\vth_{1}$) 
and population~$2$ is in a balanced two-cluster configuration (half of the oscillators in phase~$\vth_{2}$ and the other half in phase~$\vth_{2}'$) will retain this cluster configuration as phases evolve according to Eq.~\eqref{eq:OscPopulations}.
This means that the set of such states,
\[{\cal S}:= \sset{\theta_1^k\!=\!\theta_1^{n+k}\!=\!\vth_{1}, \theta_2^k\!=\!\vth_{2}, \theta_2^{n+k}\!=\!\vth_{2}', k\!=\!1,\dotsc,n},\]
is invariant under time evolution.

In the following, we consider the cluster dynamics on~${\cal S}$, which can be written in terms of two phase differences: $\psi := \vth_{2}-\vth_{1}$ and $\psi' := \vth_{2}'-\vth_{1}$.
We introduce 
 $\dom_{12}:=\omega_1-\omega_2$
to quantify the detuning of the intrinsic frequencies between the populations. From Eq.~\eqref{eq:OscPopulations}, it follows that
\begin{subequations}\label{eq:OscPopulationsRestr}
\begin{align}
\begin{split}
\dot\psi &= -\dom_{12} -  g(0)+g(\psi'-\psi)
\twocol{\\&\qquad}
+\eps\left[2g(-\psi) - g(\psi) - g(\psi')\right],
\end{split}\\
\begin{split}
\dot\psi' &= -\dom_{12} -  g(0) + g(\psi-\psi')
\twocol{\\&\qquad}
+\eps \left[2g(-\psi') - g(\psi) - g(\psi')\right],
\end{split}
\end{align}
\end{subequations}
which describes the cluster dynamics in the co-rotating reference frame of population 1.

Identifying  the initial conditions~$\theta_{(0)}\in{\cal S}$ with the corresponding initial phase differences $\psi_{(0)}:=\big(\psi(0), \psi'(0)\big)$,
we have $\Delta\Omega_{21}(\psi_{(0)}) := \Delta\Omega_{21}(\theta_{(0)}) = \lim_{T\to\infty}\frac{1}{T}\psi(T)$.

If the two populations have identical intrinsic frequencies (i.e., $\dom_{12}=0$), then, according to Eq.~\eqref{eq:OscPopulationsRestr},  states in which all oscillators are in-phase (i.e., $\psi=\psi'=0$) are stationary solutions of the dynamics. Consequently, $\Delta\Omega_{21}(0,0)=0$, and the populations are frequency synchronized. By contrast, if the intrinsic frequencies of the two populations are nonidentical (i.e., $\dom_{12}\neq0$), in general no in-phase synchronized equilibria exist. The question then is whether frequency synchronization can be achieved between two populations with nonidentical intrinsic frequencies.

To answer this question we consider the coupling function 
\begin{equation}
g(\phi) = \sin(\phi+\alpha) - r\sin(2\phi+2\alpha),
\label{eq:gfunction}
\end{equation}
 where $r>0$ controls the strength of the second harmonic and $\alpha$ is a phase frustration parameter. 
\cb{Except when noted otherwise,}  
we set 
$\alpha = -\frac{\pi}{2}$, 
which allows 
different cluster synchronous states, including weak chimeras, to be stable~\cite{bick2017robust}.
Two results follow.

\begin{figure}[t]
\centering
\includegraphics[width=0.9\linewidth]{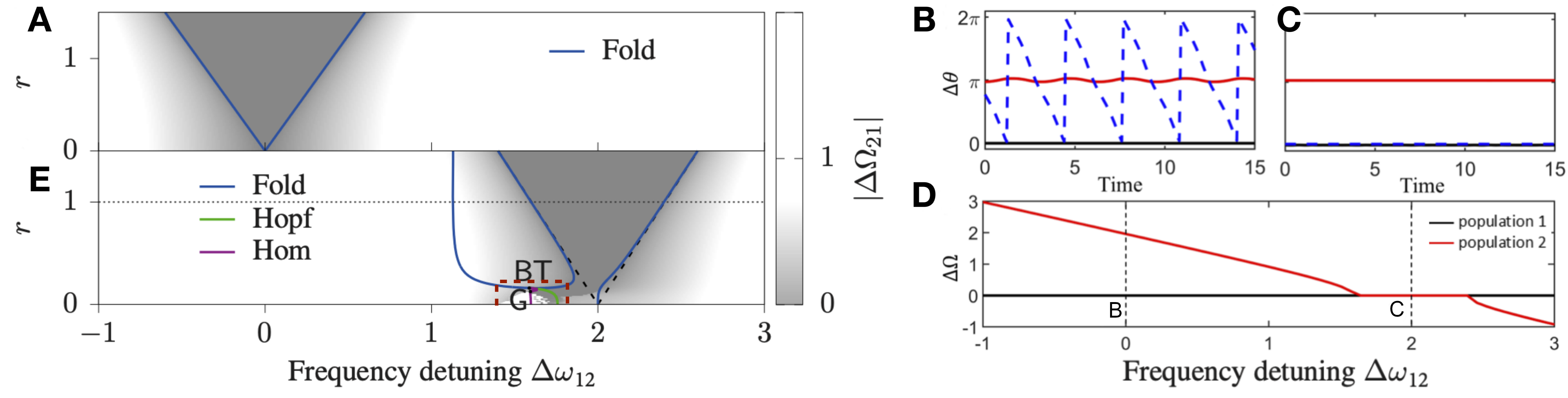}
\caption{\textbf{ Emergence of frequency synchronization in detuned phase oscillator networks. }
(\textbf{A})~Bifurcation diagram for  synchronization in the~$\dom_{12}$ vs.~$r$ space for 
the initial condition 
$\psi_{(0)}=(0.1, 0.1)$:
condition in Eq.~\eqref{eq:FreqSyncSS} 
(indistinguishable from blue lines), 
AUTO numerical analysis (blue lines), 
and direct simulations (shading).
(\textbf{B})~Time series of the phase differences for a weak chimera when $\dom_{12}=0$, where the curves indicate  
$\Delta \theta =\theta^k_{\sigma}-\theta^j_{\kappa}$ 
 in population 1 (black line),  in population 2  (red line), and between the two populations (blue dashed line).
(\textbf{C})~Phase differences for a globally synchronized state induced by frequency detuning $\dom_{12}=2$ (lines as in~{(B)}).
(\textbf{D})~Frequency difference between populations as the detuning parameter~$\dom_{12}$ is varied, where the dashed lines mark the points corresponding to panels (B) and~(C).
(\textbf{E})~Bifurcation diagram showing the emergence of synchronization induced by detuning for 
$\psi_{(0)}=(0, -\pi)$:
 condition in Eq.~\eqref{eq:FreqSyncDS} (dashed lines),  AUTO numerical analysis (solid lines), and direct simulations (shading), where the bifurcations marked are fold  (blue), Hopf  (green), and homoclinic  (purple). 
 \cb{For small $r$, the diagram shows a complex bifurcation structure with a region of frequency-synchronized periodic orbits, a Bogdanov-Takens bifurcation [BT, $(\Delta \omega_{12},r) \approx (1.588,0.160)$], and a global bifurcation of two homoclinic orbits [G, $(\Delta \omega_{12},r) \approx (1.594,0.126)$]. For clarity, a detailed zoom of the red dashed region is presented in Fig.~\ref{fig:AUTO_zoom}. }
The other parameters are $\eps=0.1$ in all panels,  $r=1$ in ~({B})-({D}), and $N=2$ in the simulations.}
\label{fig:Theory}
\end{figure}

First, for frequency synchronized populations with $\psi=\psi'$, the frequency synchrony is maintained for weak detuning~$\dom_{12}$ even if the populations are out of phase (i.e., $\psi\neq 0$); this can be related to classical studies of frequency entrainment of nonidentical oscillators~\cite{Adler1946}.
Let~${\cal SS}$ denote the points in~${\cal S}$ with $\psi=\psi'$. 
We have $\dOm_{21}(\psi,\psi') = 0$ for $(\psi, \psi')$ in ${\cal SS}$ if $(\psi, \psi')=(\psi^*, \psi^*)$ is an equilibrium point,
which is by Eq.~\eqref{eq:OscPopulationsRestr} equivalent to 
$\dom_{12} + 2\eps\left[g(\psi^*) - g(-\psi^*)\right]=0$.
For our specific choice of~$g$, this condition for frequency synchronization 
is given by 
$\dom_{12}= -4\eps r \sin (2\psi^*)$
 and is satisfied for some~$\psi^*$ provided that
\begin{equation}\label{eq:FreqSyncSS}
|\dom_{12}| \leq 4\eps r.
\end{equation}
The condition in Eq.~\eqref{eq:FreqSyncSS} is exact and defines an Arnold tongue-like cone in the $\dom_{12}$ vs.\  $r$ parameter space, as depicted in Fig.~\ref{fig:Theory}A. 
The boundaries of this region of frequency synchronization are given by fold (saddle-node) bifurcations.
Therefore, this shows how detuning can maintain frequency synchronization or result in frequency desynchronization.

Second, when the two populations are not frequency synchronized with each other in ${\cal S}$,  
frequency detuning can induce global frequency synchronization, as shown next.
Without detuning (i.e., $\dom_{12}=0$), the system can exhibit weak chimeras, which are characterized by the lack of frequency synchronization between the two populations (i.e., $\Delta\Omega_{21}(\theta_{(0)})\neq 0$)~\cite{Ashwin2014a,bick2017robust}.
A weak chimera is shown in Fig.~\ref{fig:Theory}B: population~1 is synchronized in-phase,  the clusters in population~2 are 
synchronized with a phase difference $\Delta\psi := \psi-\psi'\approx\pi$, and the phase differences between the populations  increases monotonically.
For $\eps = 0$ and
 $\omega_1=\omega_2=\omega$,
 the asymptotic frequency difference between the populations can be calculated explicitly.
In this case, for $\theta_{(0)}\in{\cal S}$ with $\Delta\psi=\pi$, we have 
$\Omega_{1}(\theta_{(0)}) = \omega + 2g(0)$ and  $\Omega_{2}(\theta_{(0)}) = \omega+g(0)+g(\pi)$.
This results in
$\Delta\Omega_{21}(\theta_{(0)}) = -g(0)+g(\pi) = 2 =: \Delta\Omega_{21}^\mathrm{u} > 0$
 for the frequency difference between the two {\it uncoupled} populations.

Now we consider the effect of intrinsic frequency detuning (i.e., $\dom_{12} > 0$) for nonzero coupling between the populations (i.e., $\eps > 0$). As $\dom_{12}$ is increased for fixed initial conditions, the frequency difference $\Delta\Omega_{21}$ between the populations reduces and then vanishes for a range of~$\dom_{12}$ values around~$\Delta\Omega_{21}^\mathrm{u}$. This is illustrated in Fig.~\ref{fig:Theory}C--D for $\eps =0.1$ and $r=1$, and it reproduces closely the behavior anticipated for the models in Fig.~\ref{fig:ni_int}.

To derive a theoretical condition for frequency synchronization induced by frequency detuning, we focus primarily on the regime $\eps\ll 1$ and large~$r$.
In this case, there is a weak chimera close to 
$\Delta\psi =\pi$ in ${\cal S}$, as described above.
Let ${\cal SC}$  denote the points in ${\cal S}$ with ${\Delta\psi =\pi}$.
In the same spirit as in the derivation of Eq.~\eqref{eq:FreqSyncSS}, imposing the existence of an equilibrium $(\psi,\psi')=(\psi^*,\psi^*-\pi)$ in ${\cal SC}$
 for Eq.~\eqref{eq:OscPopulationsRestr}
gives an approximate condition for frequency synchronization.
The condition for an equilibrium to exist
 is that
$\dom_{12} +g(0) - g(\pi) +\eps\left[g(\psi^*) - g(-\psi^*) +g(\psi^*-\pi) - g(-\psi^*+\pi)\right]=0$.
 For our choice of~$g$ and the assumption that~$r$ is large, an equilibrium exists provided that 
\vspace{-1mm}
\begin{equation}\label{eq:FreqSyncDS}
\abs{\dom_{12} - \Delta\Omega_{21}^\mathrm{u}} \leq 4\eps r.
\end{equation}
\vspace{-1mm}
This condition
gives a good approximation for the numerically computed fold bifurcation lines in Fig.~\ref{fig:Theory}E. Our numerical bifurcation analysis was implemented using AUTO~\cite{Doedel1981} and also shows that additional local and global
 bifurcations occur for  $r<1$. Nevertheless, the condition in Eq.~\eqref{eq:FreqSyncDS} provides a good approximation for~$r\gtrsim 1$. 

\cb{
\subsubsection*{Extension to more general interaction functions}

We have shown that the phase model in Eq.~\eqref{eq:OscPopulations} with the interaction function $g$ in Eq.~\eqref{eq:gfunction} can exhibit frequency synchronization with frequency detuning for $\alpha = -\pi/2$ (i.e., large phase frustration)
and $r\gtrsim 1$ (i.e., large second harmonic in $g$). 
However, the phenomenon is also expected for other $\alpha$ and $r$ values,
and more general $g$ with independent phase shifts for the first and second harmonics.
As shown in \emph{Materials and Methods}, an interaction function with up to two Fourier harmonics in $g$ is expected to produce
frequency synchronization with detuning whenever the one- and two-cluster states are stable, and there is a frequency difference between these two states in the absence of detuning. 
For $g$ in Eq.~\eqref{eq:gfunction} in the range $-\pi/2 \le  \alpha <-\pi/4 $, we show that the one-
and two-cluster states are stable for
\begin{equation}
r>r_c :=  -\cos(\alpha)/(2 \cos 2\alpha);
\label{eq:rcond}
\end{equation}
we also show that the frequency difference between the one- and two-cluster states is $\Delta\Omega_{21}^\mathrm{u} = -2 \sin(\alpha)$ and thus always nonzero for this range of $\alpha$. For such conditions, the width of the Arnold tongue can be approximated as 
\begin{equation}\label{eq:FreqSyncDSalpha}
\abs{\dom_{12} - \Delta\Omega_{21}^\mathrm{u}} \leq 4\eps r \cos(2 \alpha),
\end{equation}
which generalizes Eq.~\eqref{eq:FreqSyncDS} as a condition for detuning-induced frequency synchronization. 
Therefore, the effect is expected in the entire $(\alpha, r)$-domain above where the one- and two-cluster states are stable, which was confirmed with numerical simulations in Fig.~\ref{fig:ralpha}. 

Equation~\eqref{eq:rcond} also has another implication. For systems with large phase frustration, $\alpha \approx -\pi/2$, 
a second harmonic with a relatively small amplitude $r>0$ is sufficient for detuning-induced frequency synchronization.
However, when the phase frustration parameter is smaller in magnitude ($\alpha \rightarrow -\pi/4$), a large second harmonic $r$ is required for detuning-induced frequency synchronization ($r_c \rightarrow \infty$, see Fig.~\ref{fig:ralpha}).

We numerically calculate the phase interaction functions for the chemical oscillator and integrate-and-fire models in Fig.~\ref{fig:ni_int}. The results, presented in Fig.~\ref{fig:simg}, are in agreement with the theoretical prediction: for both models, the calculated  $g$ function implies stable one- and two-cluster states with non-zero frequency differences. 
In these examples, $g$ has a dominant first harmonic and weak second harmonic component, 
which is one of the shapes of $g$ predicted theoretically to lead to detuning-induced frequency synchronization.
}

\begin{figure} 
\centering
\includegraphics[width= 0.9\columnwidth]{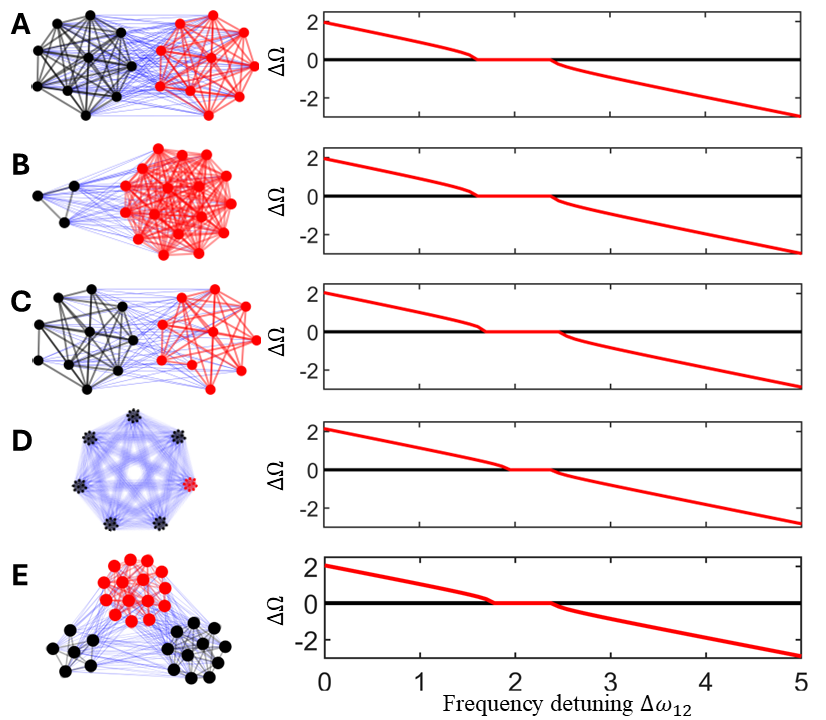} 
\caption{\cb{\textbf{ Detuning-induced frequency synchronization in increasingly complex network structures. } 
Modular networks with varying structures (left) exhibit detuning-induced frequency synchronization (right).  
(\textbf{A}) Reference case: two populations with $N=10$.  
(\textbf{B}) Two populations with 3 oscillators in population one and 17 in population two.  
(\textbf{C}) Two populations with $N=10$ and 40\% of edges randomly removed.  
(\textbf{D}) Seven populations with $N=10$.  
(\textbf{E}) Three populations with different sizes (6, 11, and 15 oscillators) and 40\% random edge removal.
Black and red indicate one- and two-cluster states, respectively. Frequencies are shown relative to the first oscillator in the first one-cluster population.  
See \emph{Materials and Methods} for model details and parameter choices. } }
 \label{fig:complex_network}
\end{figure}

\cb{
\subsubsection*{Consistent behavior across more complex networks}

Thus far, the theoretical and numerical results presented are based on all-to-all modular networks composed of two strongly coupled populations that interact through weak inter-population coupling. However, we can show that our main results extend to more general network structures, as shown in Fig.~\ref{fig:complex_network}, where we vary the number and relative size of the populations as well as the coupling density. For better comparisons across the different scenarios, the coupling is assumed to occur through mean fields 
(i.e., the strengths of the input  intra- and inter-population couplings to a node are normalized by the corresponding degrees); for further details of the model and parameter choices, we refer to \emph{Materials and Methods}. As a reference, Fig.~\ref{fig:complex_network}A shows that detuning-induced frequency synchronization readily occurs for an all-to-all modular network with two populations and $N=10$, as expected from the results above. 

We first examine the impact of varying the number of nodes within each population. As shown in Fig.~\ref{fig:complex_network}B,  frequency synchronization persists even with a strong population imbalance, where one population contains only 3 oscillators and the other 17 oscillators. This robustness arises from the mean-field nature of the interactions, which makes the dynamics largely insensitive to population size. (Fig.~\ref{fig:cn_nn} shows additional examples across different population ratios.)
A similar invariance is observed when randomly removing edges in networks with equally sized populations. As illustrated in Fig.~\ref{fig:complex_network}C, frequency synchronization remains intact even with 40\% of edges removed. (Fig.~\ref{fig:cn_edge}  confirms this behavior for different edge removal percentages.) 
As long as one- and two-cluster states form within each population, mean-field interactions dominate, preserving synchronization. However, when the network is further sparsified, some nodes begin to decouple from the main clusters, and detuning-induced synchronization becomes limited to connected subnetworks.

Figure~\ref{fig:complex_network}D illustrates the effect of increasing the number of populations: a single two-cluster population is coupled to six one-cluster populations. While detuning can still induce synchronization, the width of the Arnold tongue is noticeably reduced. This trend is further supported by Fig.~\ref{fig:cn_np}, which shows that as the number of one-cluster populations increases, the Arnold tongue narrows progressively. This reduction arises because the mean field experienced by each population becomes increasingly dominated by the one-cluster dynamics.
Finally, the phenomenon persists even when relative population size, edge density, and number of populations are varied concurrently. This is illustrated in Fig.~\ref{fig:complex_network}E for a heterogeneous network of three populations, with different sizes and 40\% of edges randomly removed, where detuning-induced frequency synchronization still emerges, highlighting the robustness of the phenomenon.

}

\subsubsection*{Experimental demonstration in  electrochemical oscillators networks}

\begin{figure}
\centering
\includegraphics[width= 0.9\columnwidth]{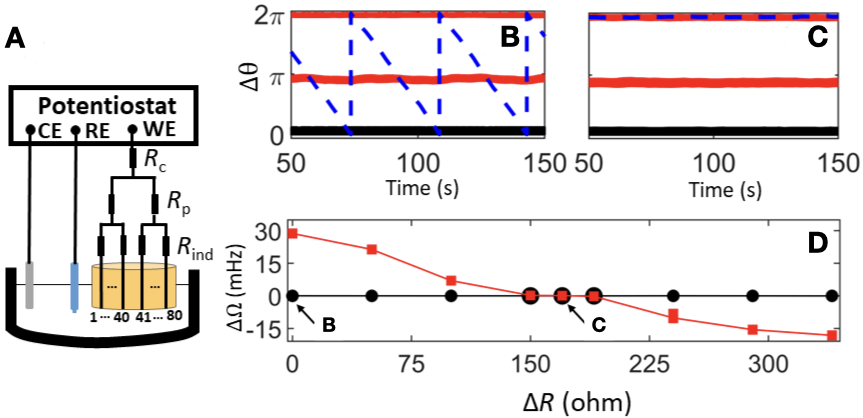}
\caption{\textbf{ Detuning-induced frequency synchronization in electrochemical oscillator experiments. }
(\textbf{A}) Experimental setup for two groups of $N=40$ oscillators, where CE, RE, and WE are the counter, reference, and working electrodes, respectively.
The oscillators are coupled through a combination of individual ($R_\textrm{ind}$), population ($R_\textrm{p}$), and global ($R_\textrm{c}$) resistances. 
(\textbf{B}) Phase differences for a desynchronized (weak chimera) state in the absence of detuning ($\Delta R =0$). 
(\textbf{C}) Phase differences for a globally synchronized state with detuning ($\Delta R = 170$ ohm). 
In panels ({B}) and ({C}), the curves indicate  $\Delta \theta =\theta^k_{\sigma}-\theta^j_{\kappa}$  in population 1 
(black line),  in population 2  (red line), and between the two populations (blue dashed line).
(\textbf{D}) Frequency 
of individual oscillators in population 1 (black) and population 2 (red) relative to 
 oscillator 1 in population 1
as a function of $\Delta R$. }
\label{fig:freq_sync}
\end{figure}

We demonstrate the emergence of frequency synchronization induced by detuning in experiments with 
oscillatory nickel electrodissolution in 3~mol/L 
H$_2$SO$_4$ electrolyte (Fig.~\ref{fig:freq_sync}A).
The nodes of our network are formed by eighty $1\,$mm-diameter wires, each connected to a potentiostat (ACM Instruments, Gill AC). 
When the potentiostat applies
 a constant circuit potential ($V = 1.260 \text{ V}$) with respect to 
a reference electrode (Hg/Hg$_2$SO$_4$ sat.~K$_2$SO$_4$), 
the rate of the metal dissolution (measured as the current) 
exhibits periodic oscillations~\cite{Kiss:2005PM}.
Applying a combination of individual, population, and global resistances 
yields a modular network structure of coupled electrochemical reactions.
The individual resistances ($R_\mathrm{ind}$) are connected directly to the electrodes and the two population 
resistances ($R_\mathrm{p}$) are connected to the individual resistances for electrodes 1--40 and 41--80 for populations~1 and~2, 
respectively. Moreover, a global resistance ($R_\mathrm{c}$) is connected to the two population resistances.
Further details on the experiment are available in the \emph{Materials and Methods} section. 

The resistors induce a strong coupling within the two populations~($K$) and a weak coupling between them ($\eps K$). 
In particular, $R_\mathrm{p}$ induces a coupling $K_\mathrm{p}=R_\mathrm{p}/[R_\mathrm{ind}(R_\mathrm{ind}+40R_\mathrm{p})]$ within each population. 
In addition, $R_\mathrm{c}$ induces a global (all-to-all)  coupling   
$K_\mathrm{c}=R_\mathrm{c}/[(R_\mathrm{ind}+40R_\mathrm{p})R_\mathrm{eq}]$, where $R_\mathrm{eq} = R_\mathrm{ind} + 40R_\mathrm{p} + 80R_\mathrm{c}$. 
Hence, the total intra- and inter-population coupling can be calculated as 
 $K=K_\mathrm{p} + K_\mathrm{c}$  and $\eps K=K_\mathrm{c}$, respectively. 

Fig.~\ref{fig:freq_sync}B shows the synchronization pattern, including phase differences, for a typical configuration ($R_\mathrm{ind} = 580 $ ohm, $R_\mathrm{p} = 9 $ ohm, and $R_\mathrm{c}$ = 0.5 ohm) with intra-population coupling $K = 17.0$ $\mu$S and cross-coupling factor $\eps = 0.03$. 
 In population 1, the oscillators are in-phase synchronized (forming a one-cluster state) with 
a frequency of 0.378 Hz. In population 2, the oscillators form a two-cluster state in which the clusters are anti-phase synchronized with a common frequency of 0.406 Hz.
Because the one- and the two-cluster populations have a frequency difference of 28 mHz, the phase difference between the populations drifts approximately linearly. 
This behavior confirms the existence of a weak chimera state \cite{Ashwin2014a, Bick2015c}.

To explore the impact of frequency detuning on the synchronization pattern, we note that the intrinsic frequencies 
of the oscillators can be varied by changing the individual resistances  ($R_\mathrm{ind}$). 
As these resistances are increased, the intrinsic frequencies of the oscillators increase at the approximately linear rate of 0.35~mHz/ohm (see  \emph{Materials and Methods}). Fig.~\ref{fig:freq_sync}C shows the synchronization pattern when $R_\mathrm{ind}$ of each wire in population~1 
is increased by $170$ ohm relative to $R_\mathrm{ind}$ in population 2 (i.e., $\Delta R = 170 $ ohm). While populations~1 and~2 retain their one- and two-cluster states, respectively, 
they now evolve at the {\it same} frequency of $0.409$~Hz. 
Therefore, the experiments confirm our prediction that detuning the intrinsic frequencies between the two populations can result 
in globally frequency-synchronized dynamics.

Fig.~\ref{fig:freq_sync}D shows the frequencies of the individual oscillators (relative to the first oscillator in population~1) for a range of values of~$\Delta R$. Without detuning (i.e., $\Delta R = 0$~ohm), there is a $28$~mHz frequency difference between the oscillators in populations~1 and~2. 
When the intrinsic frequency in population~1 is increased by increasing~$\Delta R$, the frequency difference decreases; for $\Delta R=100$~ohm, the frequency difference is~7~mHz. When $\Delta R$ is further increased, the system reaches a globally synchronized state ($\Delta \Omega_{21} = 0$~mHz) for the range $150~\textrm{ohm} \le \Delta R \le 190~\textrm{ohm}$. 
For $\Delta R > 190~\textrm{ohm}$, the frequency difference becomes negative, and the two populations are no longer synchronized with each other. 
This confirms that the detuning-induced synchronization takes place through an entrainment process similar to those resulting 
from an Arnold tongue in Fig.~\ref{fig:Theory}D.

\begin{figure}
\centering
\includegraphics[width= 0.9\columnwidth]{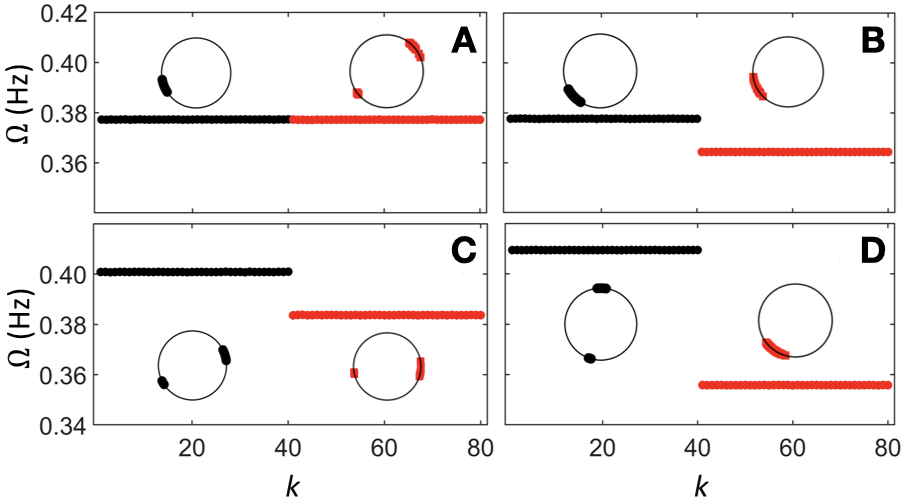}
\caption{\textbf{ Interplay between frequency detuning and initial conditions in electrochemical experiments. }
Each panel shows the frequency indexed by the oscillator number in population 1 (black, 1-40) and 2 (red, 41-80) for the parameters used in Fig.~\ref{fig:freq_sync}C.
The panels differ by their initial conditions, which lead to stable states with phases in population $\sigma$ forming either one (I) or two (II) clusters.
(\textbf{A}) Global frequency-synchronized state with a I--II phase configuration.
(\textbf{B})-(\textbf{D}) Desynchronized states with a I--I,  II--II, and  II--I configuration, respectively.
The insets show phase snapshots in each population. 
}
\label{fig:clusters}
\end{figure}

Thus far, all presented experiments had initial conditions in which populations 1 and 2 were close to one- and two-cluster states, respectively. 
Fig.~\ref{fig:clusters} presents experimental results for several different initial conditions at the  fixed detuning $\Delta R = 170~\textrm{ohm}$, 
where each main panel shows the frequency of the individual oscillators and the inset depicts a snapshot of the final state for the given initial condition.
Global frequency synchronization is observed for initial conditions close to the one-cluster state for population 1 and the two-cluster state for population 2 (Fig.~\ref{fig:clusters}A), as in Fig.~\ref{fig:freq_sync}C.
In contrast, when both populations have similar initial conditions (one cluster in Fig.~\ref{fig:clusters}B and two clusters in Fig.~\ref{fig:clusters}C), 
population~1 exhibits a higher frequency and thus global synchronization is not observed.
As shown in Fig.~\ref{fig:clusters}D, the frequency difference is even larger (53 mHz) when the initial conditions in Fig.~\ref{fig:clusters}A are swapped between the two populations.
Different initial conditions lead to qualitatively different outcomes due to multistability. 
Most importantly, the experiments reveal an interplay between frequency detuning and the initial conditions: the detuning used in  Fig.~\ref{fig:clusters}
does not yield global frequency synchronization for identical initial phases but it does for initial (and final) split phases.

\cb{
We also extracted the experimental interaction function using data corresponding to the desynchronized state shown in Fig.~\ref{fig:clusters}B. The resulting function, presented in Fig.~\ref{fig:expg}, aligns with theoretical predictions: it supports stable one- and two-cluster states with non-zero frequency differences. The experimentally obtained interaction function $g$ is dominated by a second harmonic component, with additional contributions from the first and third harmonics. These findings confirm the theoretical insight that frequency synchronization under detuning can occur even when the first harmonic is weak, provided that $g$ contains a sufficiently strong, phase-attracting second harmonic component.
}

\subsection*{Discussion}
This work shows that heterogeneity-induced frequency synchronization can be observed for a general class of oscillator networks.
This is achieved by allowing the stabilized states to be frequency synchronized without constraining them to exhibit identical phases.
This effect constitutes a generalization for synchronization \cb{in multistable modular networks} of the recently discovered phenomenon of converse symmetry breaking \cite{nishikawa2016symmetric,hart2019topological,molnar2020network,zhang2021random}, in which dynamical order is induced by system disorder.
Unlike previous work, this study shows that heterogeneity can promote global frequency synchronization of otherwise cluster-synchronized or chimera states
 even when:
(i)  the coupled systems are  one-dimensional phase oscillators, 
(ii)  the heterogeneity is implemented directly on the intrinsic frequencies, and
(iii)  the same heterogeneity is implemented for all oscillators in a network module.
The latter shows that even low-dimensional parameter manipulation can realize the full potential of detuning-induced synchronization.
We identified three key ingredients for this phenomenon to occur: coupling involving higher harmonics, phase frustration, 
and weak coupling between the modules, which are all features common to many systems. 
Indeed,  higher-order harmonics are required to describe nonlinear terms for phase reduction away from 
Hopf and  homoclinic  bifurcations  \cite{Kori:2014mz,Kori2018}, phase frustration  emerges naturally  in 
the presence of  amplitude-dependent oscillation periods and coupling delays  \cite{sakaguchi1986soluble,Kiss2007},
and real networks are known to often exhibit modular structure \cite{girvan2002community}.
As such, many biological and physical oscillator networks have the potential to exhibit frequency synchronization 
induced by frequency detuning, with important implications for their stability, optimization, and control.

\subsection*{Materials and Methods}

\cb{
\subsubsection*{Theory for more general interaction functions}

To show that frequency detuning can induce frequency synchronization for a range of phase frustration (or phase shift) parameter  
$\alpha$ of the phase model,  we consider a general phase interaction function with two harmonics given by
\begin{align}\label{eq:SupPhasInt}
g(\phi) &= \sin(\phi+\alpha) - r\sin(2\phi+\beta).
\end{align}
Here $\alpha$ and  $\beta$ are the phase shifts of the first and second harmonic, respectively, and $r$ is the amplitude of the second harmonic relative to the first. In the main text, the special case $\beta=2\alpha$ is considered. Here, we separate contributions from $\alpha$ and $\beta$ to better represent the role of the phase shifts and second harmonic amplitude. 
For $\beta\in[\frac{\pi}{2}, \frac{3\pi}{2}]$ and $r>0$,
the attracting second harmonic can 
lead to coexisting stable one- and 
two-cluster states;
explicit conditions can be stated in terms of the derivative of~$g$~\cite{Ashwin1992}.

The frequency difference~$\Delta\Omega^\text{u}_{21}$ between the phase synchronized solution $\delta\psi=0$ and the two-cluster solution $\Delta\psi=\pi$ is
\begin{align}
\Delta\Omega^\text{u}_{21} &= -g(0)+g(\pi) = -2\sin(\alpha),
\end{align}
independent of $r$ and $\beta$.
This means that the frequency difference is maximal for $\alpha = \pm\frac{\pi}{2}$ and detuning is required for frequency synchronization to emerge.

As outlined in the main text, the existence of an equilibrium $(\psi,\psi')=(\psi^*,\psi^*-\pi)$, and thus of $\psi^*$ such that $0=\Delta\omega_{12} - \Delta\Omega^\text{u}_{21} + \varepsilon[g(\psi^*)-g(-\psi^*) + g(\psi^*-\pi)- g(-\psi^*+\pi)]$, gives a condition for frequency synchronization.
For the phase interaction function~\eqref{eq:SupPhasInt}, this condition reads
\begin{align}\label{eq:SupDiff}
\Delta\omega_{12} - \Delta\Omega^\text{u}_{21} &= \varepsilon\big[2\sin(\psi^*)\cos(\alpha)-2\cos(\psi^*)\sin(\alpha)-4r\sin(2\psi^*)\cos(\beta)\big].
\end{align}
If the second harmonic is dominant (i.e., $r\gg 0$) and $\beta\approx-\pi$, we can focus on the last term to obtain the condition
\begin{align}\label{eq:SupCond}
\abs{\Delta\omega_{12} - \Delta\Omega^\text{u}_{21}} &\leq 4\varepsilon r\cos(\beta).
\end{align}
This coincides with Eq.\ \eqref{eq:FreqSyncSS} for~$\beta=-\pi$. Thus, the phase locking condition depends on both the strength and phase shift of the second harmonic.
}

\cb{
\subsubsection*{Phase model simulations for more complex networks}

The simulations in  Figs.~\ref{fig:complex_network} and \ref{fig:cn_nn}--\ref{fig:cn_np} on more general networks are performed using the following phase model: 
\begin{equation} 
\dot{\theta}^k = \omega^k + \sum_{l=1}^{N_{\textrm{tot}}} K_{kl} \, g(\theta^{l}-\theta^{k}),
\label{eq:pm_complex}
\end{equation}
where $N_{\textrm{tot}}$ is the total number of oscillators and $K_{kl}$ is the coupling strength 
between oscillators $l$ and $k$; as before, $g$ is the interaction function, and
$\theta^{k}$ and $\omega^{k}$ are respectively the phase and intrinsic frequency of oscillator $k$.
This model reduces to Eq.\ \eqref{eq:OscPopulations} in the case of two populations of $N$ globally
coupled oscillators, given suitable choices of $K_{kl}$, even though the individual populations are 
not separately indexed in this formulation

To specify $K_{kl}$, each oscillator is assigned to the respective population $\sigma$, and the coupling within and across populations is assumed to occur through a mean field. Specifically, the coupling strength between nodes $l$ and $k$ in the same population is  $K_{kl} = 2/d'_k$, where $d'_k$ is the (in-)degree of node $k$ within its population subgraph.
Similarly,  the coupling strength between nodes $l$ and $k$ in different populations is 
$K_{kl} = 2 \varepsilon /d''_k$, where $d''_k$ is the degree of node $k$ within the subgraph formed by inter-population couplings.

The intrinsic frequencies are set to $\omega^k=0$ for populations in a two-cluster state and $\omega^k=\Delta \omega_{12}$ for populations in a one-cluster state.  The detuning is thus implemented by increasing the intrinsic frequency of the oscillators in the populations with one-cluster states. All simulations are performed using  $\varepsilon = 0.1$ and $g$ as defined in Eq.~\eqref{eq:gfunction} for $r=1$ and $\alpha=-\pi/2$, where the letter leads to self-coupling as in the other networks considered in this study.
}

\subsubsection*{Experimental setup}

The experimental setup consists of a standard three-electrode cell with a counter (platinum-coated  titanium rod), reference 
(Hg/Hg$_2$SO$_4$ sat.\ K$_2$SO$_4$), and a working electrode array (nickel wires of 1 mm in diameter) connected to a potentiostat (AC Instruments Gill AC). The electrolyte was 3 mol/L H$_2$SO$_4$ held at a constant temperature of 10 $^o$C. The electrode array consists of eighty nickel wires embedded in epoxy with a spacing of 3 mm, so the electrochemical reaction only takes place on the surface end exposed to the solution. We set the circuit potential at $V$ = 1.260 ~V and measured the current with an acquisition rate of 200~Hz  using a Labview National Instruments interface. 

\subsubsection*{Effect of individual resistances on intrinsic frequencies}

We measured the effect of the individual resistances on the intrinsic frequencies of all eighty uncoupled oscillators in our experiment (Fig.\ \ref{fig:freq_sync}A). In these measurements, 
only individual resistors were used (i.e.,  $R_{\rm p}=R_{\rm c}=0$ ohm).  
The intrinsic frequencies increased approximately linearly with a slope of 0.35 mHz/ohm as the individual resistances were increased (Fig.~\ref{fig:freq_ind}).

\clearpage 

\bibliography{ref} 

\begin{thebibliography}{10}
\providecommand{\url}[1]{\texttt{#1}}
\expandafter\ifx\csname urlstyle\endcsname\relax
  \providecommand{\doi}[1]{doi:\discretionary{}{}{}#1}\else
  \providecommand{\doi}{doi:\discretionary{}{}{}\begingroup
  \urlstyle{rm}\Url}\fi

\bibitem{Pikovsky2003}
A.~Pikovsky, M.~Rosenblum, J.~Kurths, \emph{{Synchronization: A Universal
  Concept in Nonlinear Sciences}} (Cambridge University Press) (2003).

\bibitem{Strogatz2004}
S.~H. Strogatz, \emph{{Sync: The Emerging Science of Spontaneous Order}}
  (Penguin) (2004).

\bibitem{arenas2008synchronization}
A.~Arenas, A.~D{\'\i}az-Guilera, J.~Kurths, Y.~Moreno, C.~Zhou, Synchronization
  in complex networks. \emph{Phys. Rep.} \textbf{469}~(3), 93--153 (2008).

\bibitem{Panaggio2015}
M.~J. Panaggio, D.~M. Abrams, {Chimera states: coexistence of coherence and
  incoherence in networks of coupled oscillators}. \emph{Nonlinearity}
  \textbf{28}~(3), R67--R87 (2015).

\bibitem{omel2018mathematics}
O.~E. Omel'chenko, The mathematics behind chimera states. \emph{Nonlinearity}
  \textbf{31}~(5), R121 (2018).

\bibitem{wiley2006size}
D.~A. Wiley, S.~H. Strogatz, M.~Girvan, The size of the sync basin.
  \emph{Chaos} \textbf{16}~(1), 015103 (2006).

\bibitem{Martens2015}
E.~A. Martens, M.~J. Panaggio, D.~M. Abrams, {Basins of attraction for chimera
  states}. \emph{New. J. Phys.} \textbf{18}~(2), 022002 (2016).

\bibitem{Sebek:2019gy}
M.~Sebek, I.~Z. Kiss, {Plasticity facilitates pattern selection of networks of
  chemical oscillations}. \emph{Chaos} \textbf{29}~(8), 083117 (2019).

\bibitem{abrams2016introduction}
D.~M. Abrams, L.~M. Pecora, A.~E. Motter, Introduction to focus issue: Patterns
  of network synchronization. \emph{Chaos} \textbf{26}~(9), 094601 (2016).

\bibitem{nicolaou2019multifaceted}
Z.~G. Nicolaou, D.~Ero\u{g}lu, A.~E. Motter, Multifaceted dynamics of Janus
  oscillator networks. \emph{Phys. Rev. X} \textbf{9}~(1), 011017 (2019).

\bibitem{matheny2019exotic}
M.~H. Matheny, \emph{et~al.}, Exotic states in a simple network of
  nanoelectromechanical oscillators. \emph{Science} \textbf{363}~(6431),
  eaav7932 (2019).

\bibitem{Kuramoto}
Y.~Kuramoto, \emph{{Chemical Oscillations, Waves, and Turbulence}}, vol.~19 of
  \emph{Springer Series in Synergetics} (Springer) (1984).

\bibitem{Ashwin2006}
P.~Ashwin, O.~Burylko, Y.~L. Maistrenko, O.~V. Popovych, {Extreme Sensitivity
  to Detuning for Globally Coupled Phase Oscillators}. \emph{Phys. Rev. Lett.}
  \textbf{96}~(5), 054102 (2006).

\bibitem{nishikawa2016symmetric}
T.~Nishikawa, A.~E. Motter, Symmetric states requiring system asymmetry.
  \emph{Phys. Rev. Lett.} \textbf{117}, 114101 (2016).

\bibitem{zhang2021random}
Y.~Zhang, J.~L. Ocampo-Espindola, I.~Z. Kiss, A.~E. Motter, Random
  heterogeneity outperforms design in network synchronization. \emph{Proc.
  Natl. Acad. Sci. U.S.A.} \textbf{118}~(21), e2024299118 (2021).

\bibitem{barahona2002synchronization}
M.~Barahona, L.~M. Pecora, Synchronization in small-world systems. \emph{Phys.
  Rev. Lett.} \textbf{89}~(5), 054101 (2002).

\bibitem{yu2009pinning}
W.~Yu, G.~Chen, J.~L{\"u}, On pinning synchronization of complex dynamical
  networks. \emph{Automatica} \textbf{45}~(2), 429--435 (2009).

\bibitem{motter2005network}
A.~E. Motter, C.~Zhou, J.~Kurths, Network synchronization, diffusion, and the
  paradox of heterogeneity. \emph{Phys. Rev. E} \textbf{71}~(1), 016116 (2005).

\bibitem{perego2022synchronization}
A.~M. Perego, Synchronization and amplification enabled by diversity in
  nonlinear optical systems and the analogy with converse symmetry breaking for
  coupled oscillators. \emph{Phys. Rev. A} \textbf{106}~(3), L031505 (2022).

\bibitem{yang2022emergent}
J.~F. Yang, \emph{et~al.}, Emergent microrobotic oscillators via
  asymmetry-induced order. \emph{Nat, Commun.} \textbf{13}~(1), 5734 (2022).

\bibitem{medeiros2021asymmetry}
E.~S. Medeiros, U.~Feudel, A.~Zakharova, Asymmetry-induced order in multilayer
  networks. \emph{Phys. Rev. E} \textbf{104}~(2), 024302 (2021).

\bibitem{bolotov2024breathing}
M.~I. Bolotov, V.~O. Munyayev, L.~A. Smirnov, G.~V. Osipov, I.~Belykh,
  Breathing and switching cyclops states in Kuramoto networks with higher-mode
  coupling. \emph{Phys. Rev. E} \textbf{109}~(5), 054202 (2024).

\bibitem{garbin2020asymmetric}
B.~Garbin, \emph{et~al.}, Asymmetric balance in symmetry breaking. \emph{Phys.
  Rev. Res.} \textbf{2}~(2), 023244 (2020).

\bibitem{nair2021using}
N.~Nair, K.~Hu, M.~Berrill, K.~Wiesenfeld, Y.~Braiman, Using disorder to
  overcome disorder: A mechanism for frequency and phase synchronization of
  diode laser arrays. \emph{Phys. Rev. Lett.} \textbf{127}~(17), 173901 (2021).

\bibitem{gast2024neural}
R.~Gast, S.~A. Solla, A.~Kennedy, Neural heterogeneity controls computations in
  spiking neural networks. \emph{Proc. Natl. Acad. Sci. USA} \textbf{121}~(3),
  e2311885121 (2024).

\bibitem{lorch2017quantum}
N.~L{\"o}rch, S.~E. Nigg, A.~Nunnenkamp, R.~P. Tiwari, C.~Bruder, Quantum
  synchronization blockade: Energy quantization hinders synchronization of
  identical oscillators. \emph{Phys. Rev. Lett.} \textbf{118}~(24), 243602
  (2017).

\bibitem{Acebron2005}
J.~Acebr{\'{o}}n, L.~Bonilla, C.~{P{\'{e}}rez Vicente}, F.~Ritort, R.~Spigler,
  {The Kuramoto model: A simple paradigm for synchronization phenomena}.
  \emph{Rev. Mod. Phys.} \textbf{77}~(1), 137--185 (2005).

\bibitem{Wiesenfeld1998}
K.~Wiesenfeld, P.~Colet, S.~H. Strogatz, {Frequency locking in Josephson
  arrays: Connection with the Kuramoto model}. \emph{Phys. Rev. E}
  \textbf{57}~(2), 1563--1569 (1998).

\bibitem{doi:10.1126/sciadv.abb2637}
D.~Călugăru, J.~F. Totz, E.~A. Martens, H.~Engel, First-order synchronization
  transition in a large population of strongly coupled relaxation oscillators.
  \emph{Sci. Adv.} \textbf{6}~(39), eabb2637 (2020).

\bibitem{abrams2008solvable}
D.~M. Abrams, R.~Mirollo, S.~H. Strogatz, D.~A. Wiley, Solvable model for
  chimera states of coupled oscillators. \emph{Phys. Rev. Lett.}
  \textbf{101}~(8), 084103 (2008).

\bibitem{tinsley2012chimera}
M.~R. Tinsley, S.~Nkomo, K.~Showalter, Chimera and phase-cluster states in
  populations of coupled chemical oscillators. \emph{Nat. Phys.}
  \textbf{8}~(9), 662 (2012).

\bibitem{martens2013chimera}
E.~A. Martens, S.~Thutupalli, A.~Fourri{\`e}re, O.~Hallatschek, Chimera states
  in mechanical oscillator networks. \emph{Proc. Natl. Acad. Sci. U.S.A.}
  \textbf{110}~(26), 10563--10567 (2013).

\bibitem{panaggio2016chimera}
M.~J. Panaggio, D.~M. Abrams, P.~Ashwin, C.~R. Laing, Chimera states in
  networks of phase oscillators: The case of two small populations. \emph{Phys.
  Rev. E} \textbf{93}~(1), 012218 (2016).

\bibitem{bick2017robust}
C.~Bick, M.~Sebek, I.~Z. Kiss, Robust weak chimeras in oscillator networks with
  delayed linear and quadratic interactions. \emph{Phys. Rev. Lett.}
  \textbf{119}~(16), 168301 (2017).

\bibitem{Kiss:2005PM}
I.~Z. Kiss, Y.~Zhai, J.~L. Hudson, {Predicting mutual entrainment of
  oscillators with experiment-based phase models.} \emph{Phys. Rev. Lett.}
  \textbf{94}~(24), 248301 (2005).

\bibitem{Kori2018}
H.~Kori, I.~Z. Kiss, S.~Jain, J.~L. Hudson, Partial synchronization of
  relaxation oscillators with repulsive coupling in autocatalytic
  integrate-and-fire model and electrochemical experiments. \emph{Chaos}
  \textbf{28}~(4), 045111 (2018).

\bibitem{Ashwin2014a}
P.~Ashwin, O.~Burylko, {Weak chimeras in minimal networks of coupled phase
  oscillators}. \emph{Chaos} \textbf{25}, 013106 (2015).

\bibitem{Bick2015c}
C.~Bick, P.~Ashwin, {Chaotic weak chimeras and their persistence in coupled
  populations of phase oscillators}. \emph{Nonlinearity} \textbf{29}~(5),
  1468--1486 (2016).

\bibitem{strogatz2000kuramoto}
S.~H. Strogatz, From Kuramoto to Crawford: exploring the onset of
  synchronization in populations of coupled oscillators. \emph{Physica D}
  \textbf{143}~(1), 1--20 (2000).

\bibitem{Ashwin1992}
P.~Ashwin, J.~W. Swift, {The dynamics of n weakly coupled identical
  oscillators}. \emph{J. Nonlinear Sci.} \textbf{2}~(1), 69--108 (1992).

\bibitem{Golubitsky2002}
M.~Golubitsky, I.~Stewart, \emph{{The Symmetry Perspective}}, vol. 200 of
  \emph{Progress in Mathematics} (Birkh{\"{a}}user Verlag) (2002).

\bibitem{Bick2015d}
C.~Bick, {Isotropy of Angular Frequencies and Weak Chimeras with Broken
  Symmetry}. \emph{J. Nonlinear Sci.} \textbf{27}~(2), 605--626 (2017).

\bibitem{Adler1946}
R.~Adler, {A Study of Locking Phenomena in Oscillators}. \emph{Proceedings of
  the IRE} \textbf{34}~(6), 351--357 (1946).

\bibitem{Doedel1981}
E.~J. Doedel, {AUTO: A Program for the Automatic Bifurcation Analysis of
  Autonomous Systems}. \emph{Congressus Numerantium} \textbf{30}, 265--384
  (1981).

\bibitem{hart2019topological}
J.~D. Hart, Y.~Zhang, R.~Roy, A.~E. Motter, Topological control of
  synchronization patterns: Trading symmetry for stability. \emph{Phys. Rev.
  Lett.} \textbf{122}~(5), 058301 (2019).

\bibitem{molnar2020network}
F.~Molnar, T.~Nishikawa, A.~E. Motter, Network experiment demonstrates converse
  symmetry breaking. \emph{Nat. Phys.} \textbf{16}~(3), 351--356 (2020).

\bibitem{Kori:2014mz}
H.~Kori, Y.~Kuramoto, S.~Jain, I.~Z. Kiss, J.~L. Hudson, {Clustering in
  globally coupled oscillators near a Hopf bifurcation: Theory and
  experiments}. \emph{Phys. Rev. E} \textbf{89}~(6), 062906 (2014).

\bibitem{sakaguchi1986soluble}
H.~Sakaguchi, Y.~Kuramoto, A soluble active rotater model showing phase
  transitions via mutual entertainment. \emph{Prog. Theor. Phys.}
  \textbf{76}~(3), 576--581 (1986).

\bibitem{Kiss2007}
I.~Z. Kiss, C.~G. Rusin, H.~Kori, J.~L. Hudson, {Engineering Complex Dynamical
  Structures: Sequential Patterns and Desynchronization}. \emph{Science}
  \textbf{316}~(5833), 1886--1889 (2007).

\bibitem{girvan2002community}
M.~Girvan, M.~E. Newman, Community structure in social and biological networks.
  \emph{Proc. Natl. Acad. Sci. U.S.A.} \textbf{99}~(12), 7821--7826 (2002).

\bibitem{Haim1992}
D.~Haim, O.~Lev, L.~M. Pismen, M.~Sheintuch, Modeling periodic and chaotic
  dynamics in anodic nickel dissolution. \emph{J. Phys. Chem.} \textbf{96}~(6),
  2676--2681 (1992).

\bibitem{Kiss1999}
I.~Z. Kiss, W.~Wang, J.~L. Hudson, Experiments on Arrays of Globally Coupled
  Periodic Electrochemical Oscillators. \emph{J. Phys. Chem.}
  \textbf{103}~(51), 11433--11444 (1999).

\bibitem{Kiss2005}
I.~Z. Kiss, Y.~Zhai, J.~L. Hudson, Predicting Mutual Entrainment of Oscillators
  with Experiment-Based Phase Models. \emph{Phys. Rev. Lett.} \textbf{94}~(24),
  248301 (2005).

\end{thebibliography}
\bibliographystyle{sciencemag}

\paragraph*{Funding:}
The authors acknowledge support from NSF Grant No.\ CHE-1900011 (I.Z.K.); NSF Grant No. DMS-2308341, ONR Grant No.\ N00014-22-1-2200, and ARO Grant No.\ W911NF-19-1-0383 (A.E.M.);  and  CONACYT (J.L.O.-E.).
\paragraph*{Author contributions:}
\cb{JLOE: Writing - original draft, Conceptualization, Investigation, Writing - review \& editing, Methodology, Data curation, Validation, Formal analysis, Software, Visualization. CB: Writing - original draft, Conceptualization, Investigation, Writing - review \& editing, Methodology, Formal analysis, Visualization. AEM: Writing - original draft, Conceptualization, Writing - review \& editing, Funding acquisition, Supervision, Project administration. IZK: Writing - original draft, Conceptualization, Investigation, Writing - review \& editing, Methodology, Funding acquisition, Supervision,
Software, Project administration, Visualization. }
\paragraph*{Competing interests:}
The authors declare that they have no competing interests.
\paragraph*{Data and materials availability:}
\cb{The experimental data used in the paper are deposited in our Zenodo repository \url{ https://doi.org/10.5281/zenodo.15122129}. }
\cb{All other data needed to evaluate the conclusions of the paper are present in the paper and/or the Supplementary Materials}.

\subsection*{Supplementary Materials}
Supplementary Text \\
Figures S1 to S11\\

\newpage

\renewcommand{\thefigure}{S\arabic{figure}}
\renewcommand{\thetable}{S\arabic{table}}
\renewcommand{\theequation}{S\arabic{equation}}
\renewcommand{\thepage}{S\arabic{page}}
\setcounter{figure}{0}
\setcounter{table}{0}
\setcounter{equation}{0}
\setcounter{page}{1}

\begin{center}
\section*{Supplementary Materials for\\ \scititle}

	Jorge Luis Ocampo-Espindola,
	Christian Bick,
	Adilson E. Motter, and
	Istv\'{a}n Z. Kiss\\

\end{center}

\subsubsection*{This PDF file includes:}
Supplementary Text \\
Figures S1 to S11\\

\newpage

\subsection*{Supplementary Text}

\subsubsection*{Simulations with the nickel electrodissolution model}

We used a nickel electrodissolution model proposed by Haim {\it et al.} \cite{Haim1992} under potentiostatic conditions \cite{Kiss1999, Kiss2005}  to simulate detuning-induced synchronization in Fig.\ \ref{fig:ni_int}B. The behavior of two populations of 2 oscillators was simulated with a strong electric coupling within the populations ($K$) and a weak coupling ($\varepsilon K$) between the populations. Each oscillator is described by two variables,  the electrode potential $e$ and the total surface coverage of nickel oxide and hydroxide $\theta$. The two variables, $e$ and $\theta$, are governed by the following equations:

\begin{equation}
\label{EP1-equ}
\frac{de_l}{dt}=\frac{V-e_l}{R_1}-J_F(e_l, \theta_l)+K\sum\limits_{j=1}^2 (e_j-e_l)+\varepsilon K\sum\limits_{j=3}^4 (e_j-e_l)  \;\;\; \mbox{for\;\;} \textit{l}=1, 2,
\end{equation}

\begin{equation}
\label{EP2-equ}
\frac{de_l}{dt}=\frac{V-e_l}{R_2}-J_F(e_l, \theta_l)+\varepsilon K\sum\limits_{j=1}^2 (e_j-e_l)+K\sum\limits_{j=3}^4 (e_j-e_l)  \;\;\; \mbox{for\;\;} \textit{l}=3, 4,
\end{equation}

\begin{equation}
\label{theta-equ}
\Gamma \frac{d\theta_l}{dt}=\frac{\exp(0.5e_l)}{1+C_h\exp(e_l)}(1-\theta_l)-\frac{b\, C_h\exp(2e_l)\theta_l}{c\, C_h+\exp(e_l)}  \;\;\; \mbox{for\;\;}  \textit{l}=1, ... 4,
\end{equation}

where oscillators $l=(1,2)$ and $(3,4)$ belong to populations 1 and 2, respectively,  $V = 25.18$ is the dimensionless circuit potential, $R_1$ and $R_2$ are the dimensionless (equivalent) resistances of populations 1 and 2, respectively, and $t$ is the dimensionless time. Here, $\Gamma = 0.01$ is the surface capacity, $J_F(e, \theta)$ is the Faraday current density

\begin{equation}
\label{Fc-equ}
J_F(e, \theta)=\Big[\frac{C_h\exp(0.5e)}{1+C_h\exp(e)}+a\exp(e)\Big](1-\theta),
\end{equation}
and $C_h =1600$, $a=0.3$, $b=6\times10^{-5}$, and $c=1\times 10^{-3}$ are kinetic parameters.

With the resistances for the two populations set to  $R_1=R_2=20.00$ (i.e., without detuning), the system exhibits a weak chimera state. In this state,  one population remains in-phase while the oscillators in the other population are in anti-phase, and there is a nonzero frequency difference between the two populations (Fig.~\ref{fig:nickel_model}A). 
Fig.~\ref{fig:nickel_model}B shows the behavior with frequency detuning, in this case for $R_1=20.23$ while keeping $R_2=20.00$, the system exhibits global frequency synchronization in which the oscillators in population 1 are in-phase synchronized and those in population 2 are anti-phase synchronized (Fig.~\ref{fig:nickel_model}B).

Fig.\ \ref{fig:ni_int}B  shows the frequencies of the oscillators  by changing the individual resistance $R_1$ from $20.00$ 
to $20.29$ in small steps. The amount of heterogeneity was defined by the resistance change  $\Delta R=R_1-R_2$ (the sign of the heterogeneity change was chosen to be positive for the conditions considered in Fig.\ \ref{fig:ni_int}B). For $\Delta R=0$, there is a chimera state with a positive frequency difference between the populations ($18.6\times 10^{-4}$). When  $\Delta R$
is increased, the frequency difference decreases. Frequency synchronization was achieved for the region  $0.22\le \Delta R\le 0.24$. 
Further increasing  $\Delta R$ causes the two populations to desynchronize
relative to each other. 

As shown in Fig.~\ref{fig:freq_IC}A, Fig.\ \ref{fig:ni_int}B concerns initial conditions in which populations 1 and 2 are close to one- and two-cluster states, respectively. 
The impact of the same frequency detuning for several different initial conditions is shown in  Fig.~\ref{fig:freq_IC}B-D. This is consistent with the experimental results in Fig.\ \ref{fig:clusters}.

\cb{Additional simulations were performed to numerically estimate the phase interaction function $g$ in Eq.~\eqref{eq:OscPopulations} that corresponds to the dynamics of system \eqref{EP1-equ}--\eqref{theta-equ} with weak coupling. For this purpose, the simulations were performed with initial conditions close to the I-I configuration state (one cluster in each population) for $N=2$,  $\Delta R=0.03$, and the other parameters as specified earlier. The interaction function is obtained by representing the dynamics of each population by the average over the corresponding $e_l$ and plotting the instantaneous frequency of population $\sigma$ (corrected by the natural frequency)  $\dot{\theta}_\sigma - \omega_\sigma$  versus $\Delta \theta$. The resulting interaction function is shown in Fig.~\ref{fig:simg}A--B along with the magnitude of its numerically determined Fourier harmonics up to 4th order.}

\subsubsection*{Simulations with the integrate-and-fire model}
We used an autocatalytic integrate-and-fire model \cite{Kori2018} to simulate detuning-induced synchronization in Fig.\ \ref{fig:ni_int}C. The model consists of a population of four oscillators with a state variable $u_l$, for $l=1,2,3,4$, and a parameter  $p_l$  for each oscillator that determines whether
the variable is increasing or decreasing. The oscillators are governed by the integrate-and-fire equations
\begin{equation}
\label{IF-equ}
\tau_l\frac{du_l}{dt} = p_{l}u_{l} - (1-p_{l})u_{l}B+p_{l}K(\mu_{l}-u_{l}),
\end{equation}
where $l=1,2$ and $3,4$ correspond to populations 1 and 2, respectively, and $K$ is the coupling parameter.
When the variable $u_l$ reaches 1 from below, $p_l$ is set to 0, and the variable starts to decrease; when the variable $u_l$ reaches $A$ from above, $p_l$ is set to $1$, and the 
variable starts to increase. The parameters $A$, $B$, and $\tau$ describe oscillator properties, where  $\tau_l$ defines the timescale of oscillator $l$.

The variables are coupled to the corresponding mean fields as $\mu_1=\mu_2= \frac{u_1+u_2+\gamma(u_3+u_4)}{2+2\gamma}$ and $\mu_3=\mu_4= \frac{u_3+u_4+\gamma(u_1+u_2)}{2+2\gamma}$, where $\gamma$ is the cross-coupling factor between the populations. 
Note that the oscillators have a refractory period in the sense that there is no coupling
when $u_l$ is decreasing. That is, the coupling is  present only in the excitatory period (i.e., when $u_l$ is increasing). 
This model has proved useful in describing the dynamics of oscillators close to a saddle-loop  (i.e., homoclinic) bifurcation \cite{Kori2018}.

\begin{figure}
\centering
\includegraphics[width= 0.7\columnwidth]{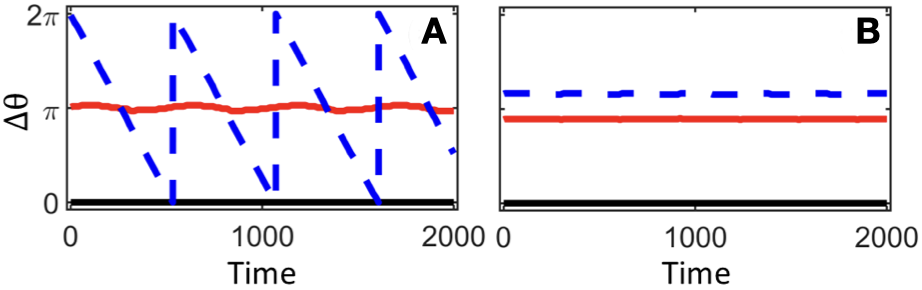}
\caption{\textbf{ Detuning-induced synchronization in the nickel electrodissolution model. }
(\textbf{A}) Time series of the phase differences for a weak chimera (frequency-desynchronized) state in the absence of detuning ($\Delta R =0$). 
(\textbf{B}) Phase difference for a globally synchronized state in the presence of detuning ($\Delta R = 0.23$). 
In both panels, the curves indicate the phase difference $\Delta \theta =\theta^k_{\sigma}-\theta^j_{\kappa}$  in population 1 
(black line),  in population 2  (red  line), and between the two populations 
(blue dashed line).
The other parameters are $N=2$, $K=0.002$, and $\varepsilon=0.05$.
}
\label{fig:nickel_model}
\end{figure}

\begin{figure} 
\centering
\includegraphics[width= 1.0\columnwidth]{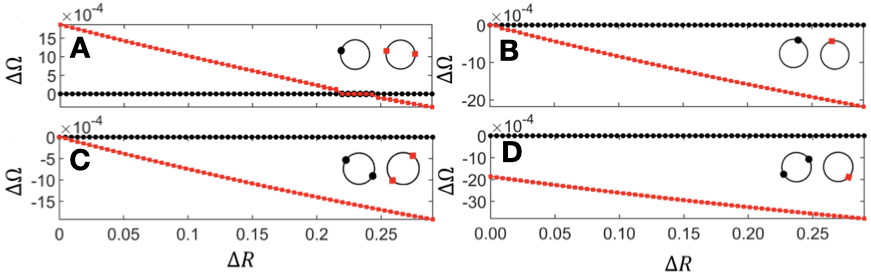}
\caption{\textbf{ Interplay between frequency detuning and initial conditions in the nickel electrodissolution model. } The oscillator frequencies in population 1 (black) and population 2 (red) (relative to oscillator 1 in population 1) are shown as functions of $\Delta$$R$ for different initial conditions. 
The insets depict phase snapshots in each population, showing the convergence to either one (I) or two (II) clusters in each population. 
(\textbf{A}) Detuning-induced synchronization for a I--II phase configuration (as considered in Fig.\ \ref{fig:ni_int}B).
(\textbf{B})-(\textbf{D})  Detuning-induced desynchronization for I--I,  II--II, and  II--I configurations, respectively.
The parameters are the same as in Fig.~\ref{fig:nickel_model}.
}
\label{fig:freq_IC}
\end{figure}

\begin{figure}
\centering
\includegraphics[width= 0.7\columnwidth]{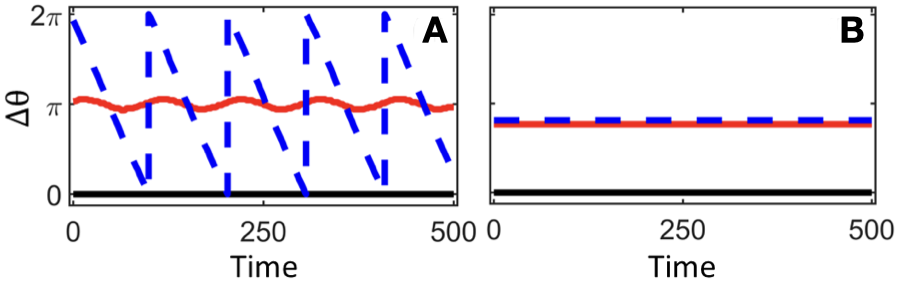}
\caption{\textbf{ Detuning-induced synchronization in the integrate-and-fire model. }
(\textbf{A}) Time series of the phase differences for a weak chimera state in the absence of detuning ($\Delta \tau =0$). 
(\textbf{B}) Phase difference for a globally synchronized state in the presence of detuning ($\Delta \tau = 0.016$). 
In both panels, the curves indicate the phase difference $\Delta \theta =\theta^k_{\sigma}-\theta^j_{\kappa}$  in population 1 
(black line), 
in population 2 
(red  line), and between the two populations 
(blue dashed line).
The other parameters are $N=2$, $K=0.1$, $\gamma=0.1$, $A=0.18$, and $B=17.0$.
 }
 \label{fig:IaF_model}
\end{figure}

Heterogeneities in parameter $\tau_l$ emulate frequency detuning in the system by setting different timescales (intrinsic periods) for the
oscillators. The timescales are assumed to be the same within each population, i.e., $\tau_1 = \tau_2$ and $\tau_3 = \tau_4$. 
 The heterogeneities are quantified through $\Delta\tau=\tau_{3}-\tau_1$ (the sign was once again chosen to be positive for the conditions considered in Fig.\ \ref{fig:ni_int}C).
Fig.~\ref{fig:IaF_model}A shows that without frequency detuning (i.e., $\Delta \tau=0$)  the system exhibits a weak chimera state.
As in the case of the nickel electrodissolution model (Fig.~\ref{fig:nickel_model}), the oscillators in population 1 are in-phase synchronized and those in population 2 are anti-phase synchronized, with the two populations exhibiting different frequencies. 
 
 Fig.~\ref{fig:IaF_model}B shows the corresponding behavior with frequency detuning, in this case for $\Delta \tau = 0.016$.  The system exhibits a globally synchronized state in which the oscillators in population 1 are in-phase and those in population 2 are in anti-phase, but with both populations now exhibiting the same frequency. 

Fig.\ \ref{fig:ni_int}C  shows the frequencies of the oscillators by changing $\Delta \tau$ in small steps from zero to $0.020$. 
For $\Delta \tau=0$, there is a weak chimera state with a frequency difference of $9\times 10^{-3}$ between the populations. 
When  $\Delta \tau$ is increased, the frequency difference between populations decreases and a frequency synchronized region appears for $0.15\le \Delta \tau\le 0.17$. 
Increasing  $\Delta \tau$ beyond $\Delta \tau =0.17$ leads to frequency desynchronization, which is similar to the behavior observed for the nickel electrodissolution model.

\begin{figure}
\centering
\includegraphics[width= 0.7\columnwidth]{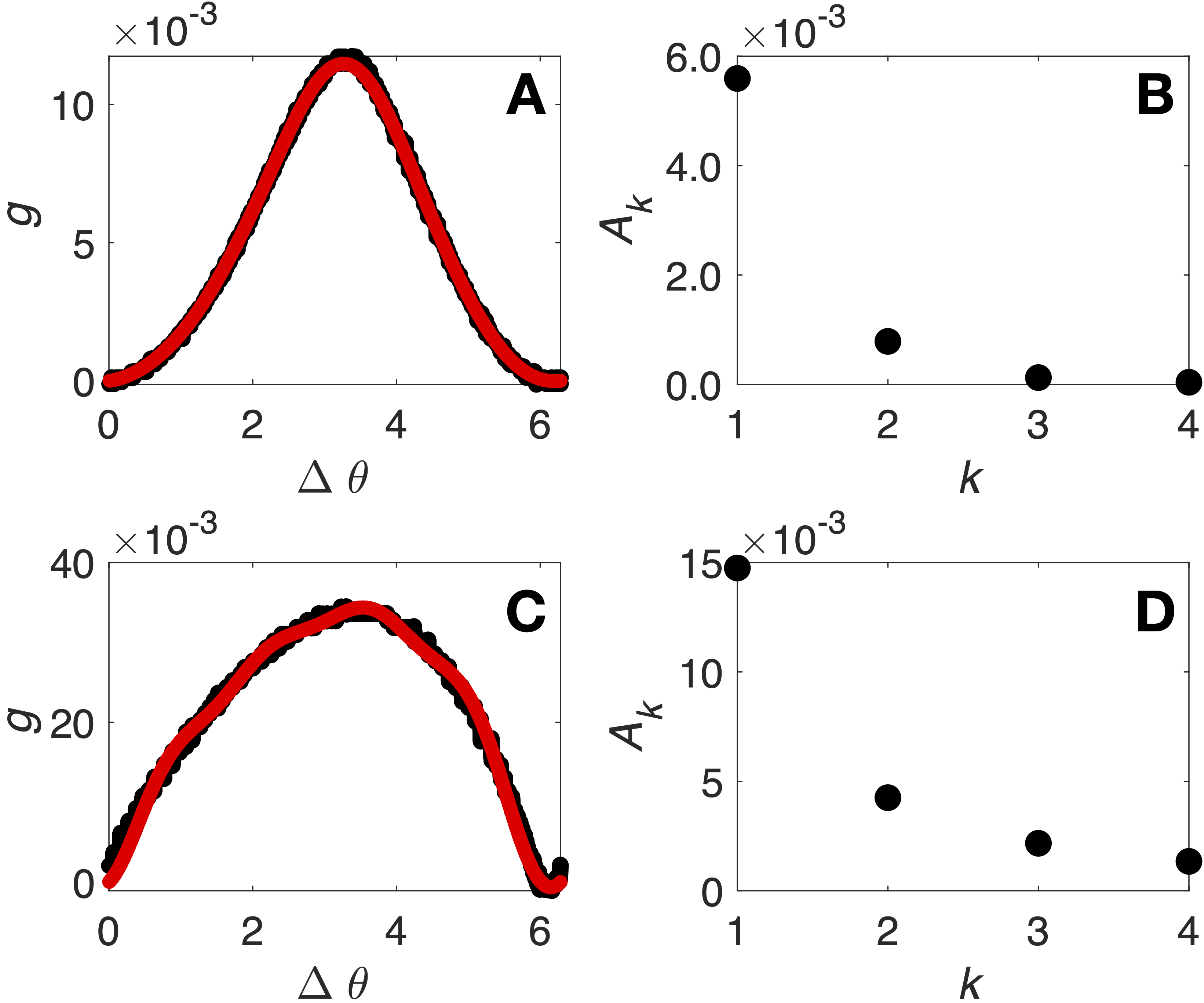}
\caption{\cb{\textbf{ Interaction functions calculated numerically for the models in Fig.\ \ref{fig:ni_int}. } 
Measured (black circles) and fitted (red curves) interaction functions (left) and the corresponding magnitudes of the first four Fourier harmonics (right). (\textbf{A}-\textbf{B}) Interaction function of the chemical oscillator model. The calculated
$g$ can be used to determine the stability of the one- and two-cluster states and their frequency difference [$\Delta \Omega_{21}^\text{u} = g(\pi) - g(0)$]. 
The obtained interaction function, which has a dominant first harmonic and small higher harmonics,  predicts a stable one-cluster state (i.e., the relevant eigenvalue is negative, $\lambda_{\max} = -2.9\times10^{-4}$) and a stable two-cluster state ($\lambda_{\max} =-0.8\times10^{-3}$)
with a frequency difference of $\Delta\Omega_{21}^\mathrm{u}=11\times10^{-3}$ between the two populations. 
(\textbf{C}-\textbf{D}) Interaction function of the integrate-and-fire model.  
Once again, the interaction function has a dominant first harmonic and predicts stable one-cluster  ($\lambda_{\max}  = -9.2\times10^{-3}$) and two-cluster $(\lambda_{\max} = -4.5\times10^{-3}$) 
states with a frequency difference of $ \Delta\Omega_{21}^\mathrm{u}=33\times10^{-3}$.
}
 }
 \label{fig:simg}
\end{figure}

\cb{We also numerically estimated the phase interaction function for the integrated-and-fire model following the same procedure used for the nickel electrodissolution model.  In this calculation, we use $N=2$, $\Delta \tau=0.1$, $K=1.3$, and the previously specified values for the other parameters. The interaction function is shown in Fig.~\ref{fig:simg}C--D along with the magnitude of the Fourier harmonics up to 4th order.}

\begin{figure}
\centering
\includegraphics[width= 0.5\columnwidth]{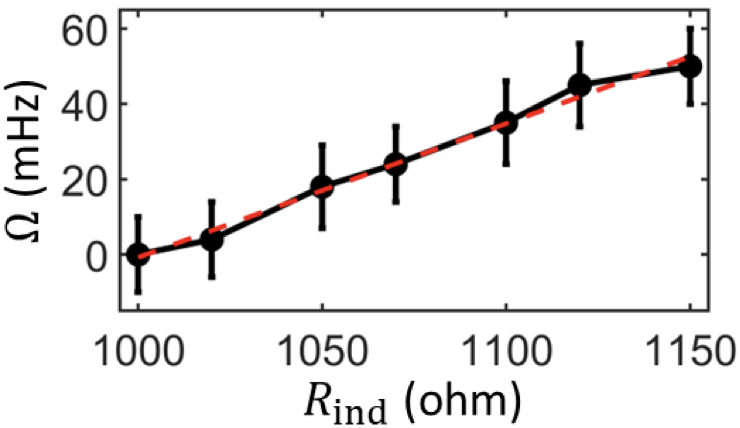}
\caption{\textbf{ The effect of individual resistances on the intrinsic frequencies of the oscillators (relative to the frequency for $R_\textrm{ind}=1000$ ohm). } The dots mark the mean values and 
the error bars represent the standard deviations for 80 nickel electrodes, which is is approximately the same for all points. The straight dashed line indicates a linear trend. 
}
\label{fig:freq_ind}
\end{figure}

\begin{figure}
\centering
\includegraphics[width= 0.7\columnwidth]{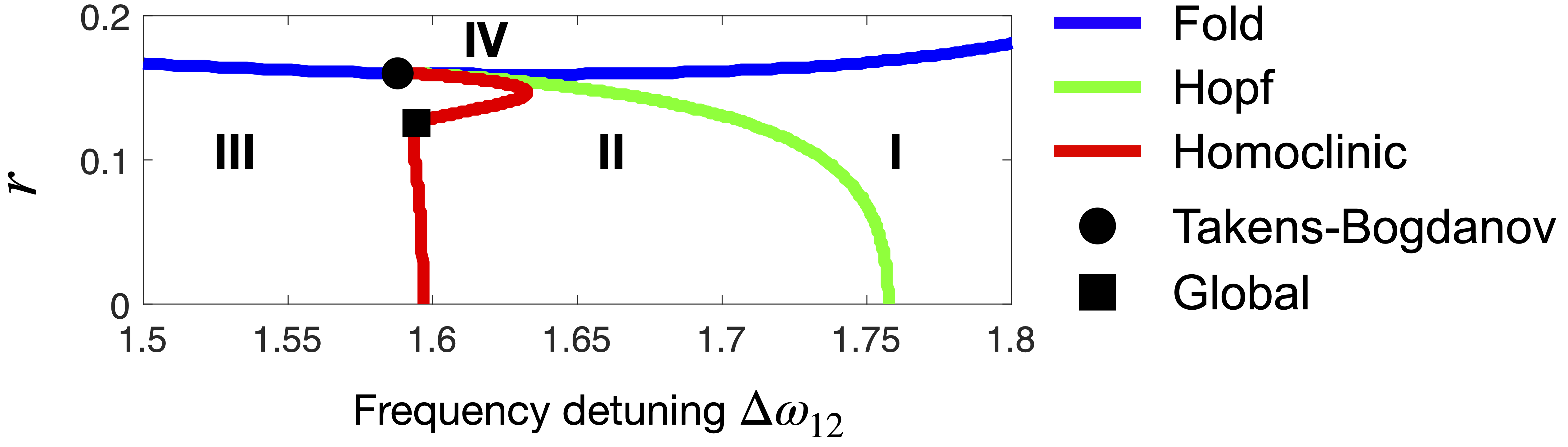}
\caption{\cb{\textbf{ Detailed bifurcation diagram in Fig.~\ref{fig:Theory}E. }
Magnified diagram of the inset in Fig.~\ref{fig:Theory}E (dashed red box). The typical qualitative 
dynamical behavior in the various regions is as follows: 
frequency synchronization with constant phase differences in both populations (region I),
frequency synchronization with zero phase differences in the one-cluster population and oscillatory phase differences in the two-cluster population (region II), and unsynchronized dynamics with zero phase differences in the one-cluster population and oscillatory phase differences in the two-cluster population (regions III and IV).}
 }
 \label{fig:AUTO_zoom}
\end{figure}

\begin{figure}
\centering
\includegraphics[width= 0.4\columnwidth]{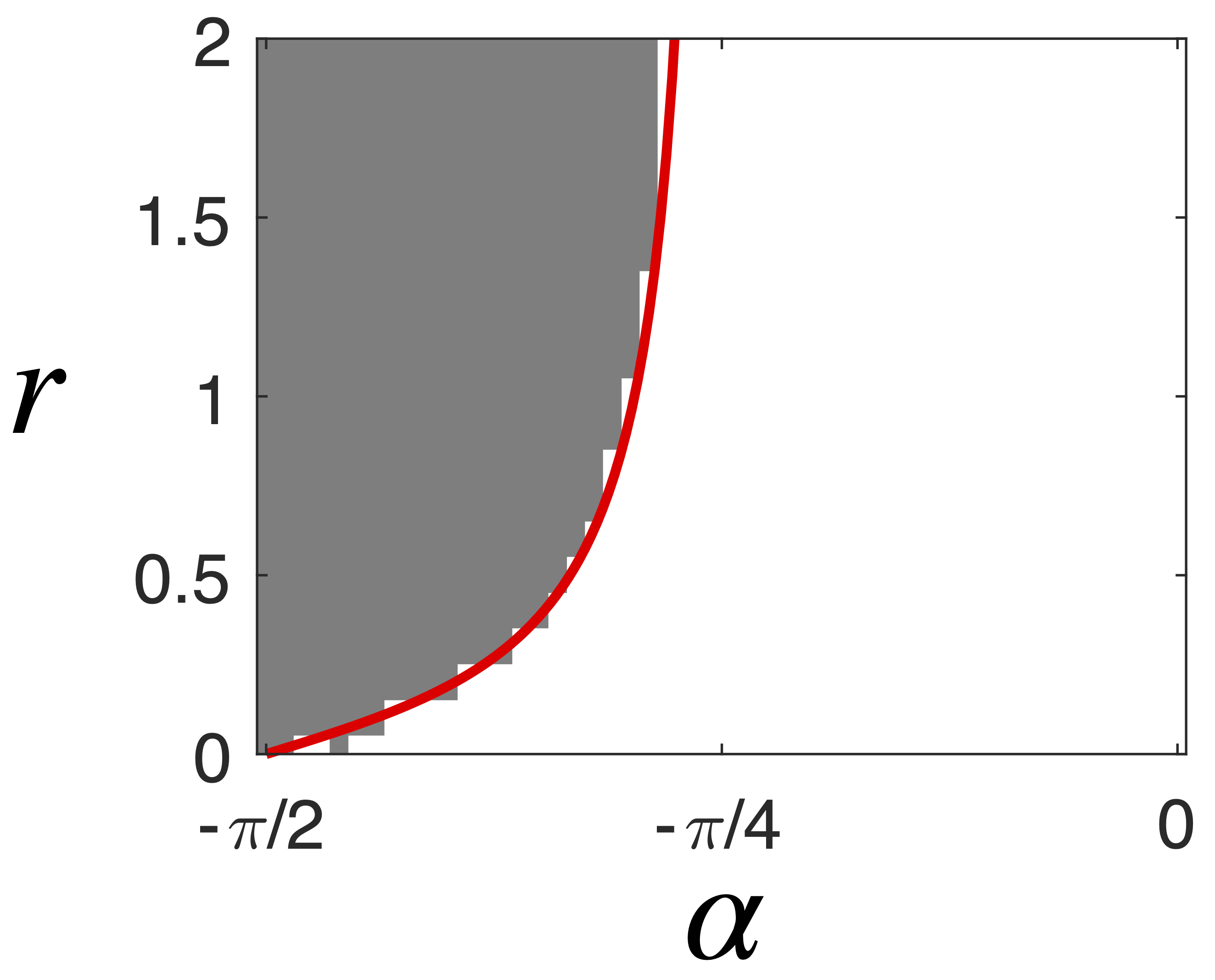}
\caption{\cb{\textbf{ Parameter region of detuning-induced frequency synchronization in the phase model. }
Phase diagram for the model in Eqs.~\eqref{eq:OscPopulations} and \eqref{eq:gfunction} with $N=2$. 
Gray region: combinations of $r$ and $\alpha$ resulting in $|\Delta \Omega_{21}| > 0$ for $\Delta \omega_{12}=0$ and  $|\Delta \Omega_{21}| =0$ for suitable $|\Delta \omega_{12}|>0$,
determined from simulations for $\varepsilon>0$.
For given $\alpha$ and $r$, the simulations were performed in the range of $\Delta \omega_{12}=[-3,3]$. 
Red curve: theoretically predicted stability boundary above which the one- and  two-cluster states are stable,
calculated for $\varepsilon=0$ as $r > -\cos(\alpha)/[2 \cos(2\alpha)]$ (see~\cite{Ashwin1992} for a derivation of the formula).  
The numerically observed and theoretically predicted domains for detuning-induced frequency synchronization show excellent agreement.}
 }
 \label{fig:ralpha}
\end{figure}

\begin{figure}
\centering
\includegraphics[width= 0.7\columnwidth]{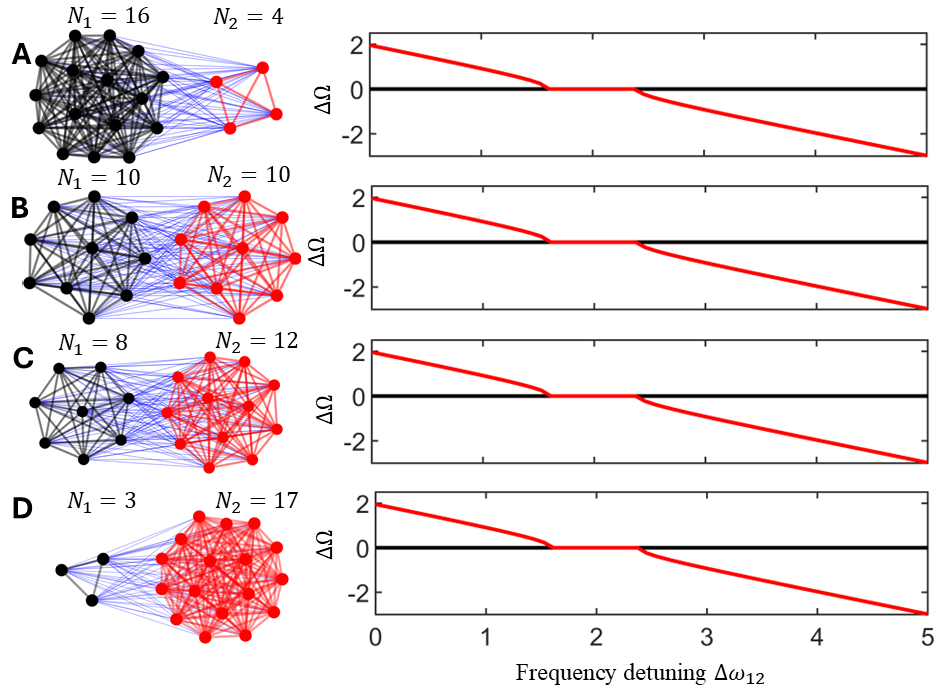}
\caption{\cb{\textbf{ Numerical simulations with phase model for populations of different sizes. } 
Networks with two all-to-all coupled populations of varying sizes (left) exhibit frequency synchronization over a similar range of frequency detunings (right). The simulations are for two-population modular networks with a population of size $N_1$ in a one-cluster state (black) and a population of size $N_2$ in a two-cluster state (red) for $N_1+N_2=20$: (\textbf{A}) $N_1=16$, $N_2=4$;  (\textbf{B}) $N_1=10$, $N_2=10$ (reference case);  (\textbf{C}) $N_1=8$, $N_2=12$; and  (\textbf{D}) $N_1=3$, $N_2=17$.
Frequencies are shown relative to the first oscillator in the one-cluster population.
For model description and other parameters, see \emph{Materials and Methods}.
}}
 \label{fig:cn_nn}
\end{figure}

\begin{figure}
\centering
\includegraphics[width= 0.7\columnwidth]{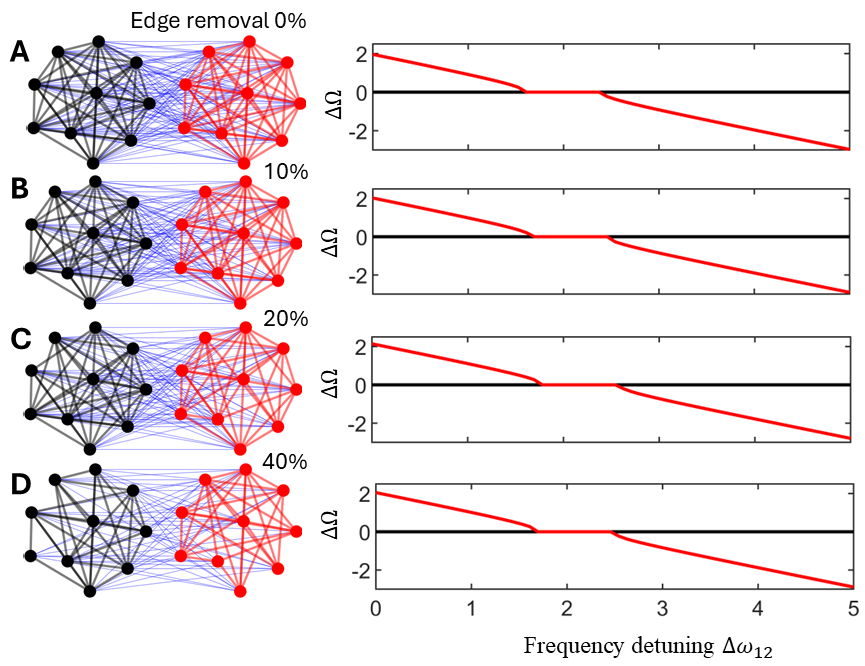}
\caption{\cb{\textbf{ Numerical simulations with phase model for random edge removal. }
Networks with different extents of random edge removal (left) exhibit detuning-induced frequency synchronization over approximately the same range of frequency detunings (right).
The simulations are for two populations of $N=10$ nodes, arranged in all-to-all modular networks:
(\textbf{A}) without edge removal (reference case), 
(\textbf{B}) with 10\% of edge removal,
(\textbf{C}) with  20\% of edge removal, and  
 (\textbf{D}) with 40\% of edge removal.
 As before, we use black for the one-cluster state, red for the two-cluster state, 
 and  frequencies relative to the first oscillator of the one-cluster population.) 
 For model description and other parameters, see \emph{Materials and Methods}.
 }}
 \label{fig:cn_edge}
\end{figure}

\begin{figure}
\centering
\includegraphics[width= 0.8\columnwidth]{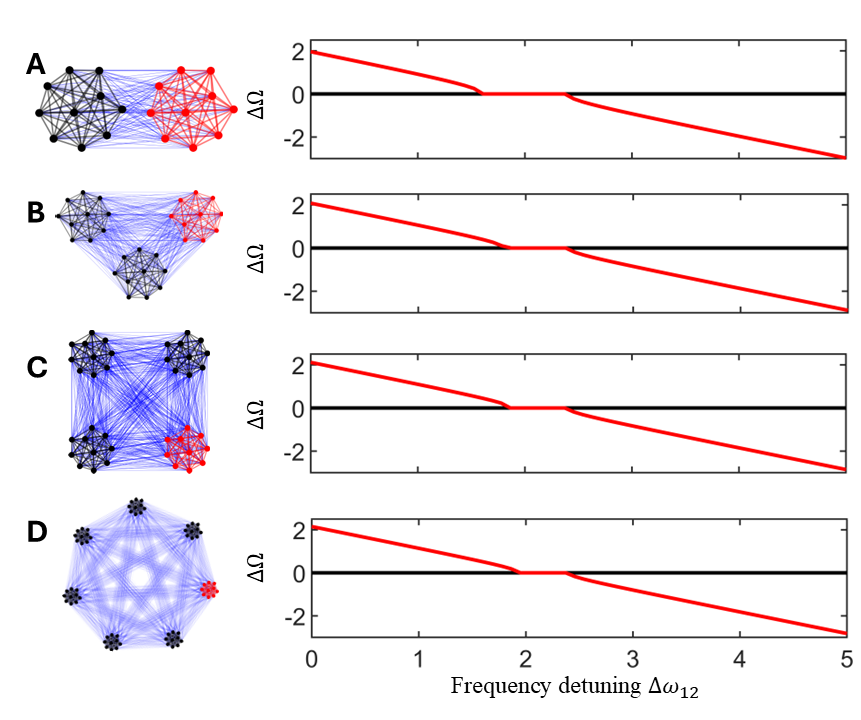}
\caption{\cb{\textbf{ Numerical simulations with phase model for a varying number of populations. }
Networks with different numbers of all-to-all coupled populations (left) exhibit detuning-induced frequency synchronization (right).
The simulations are for modular networks with 
(\textbf{A})  $N_{\textrm {pop}}=2$ populations (reference case), 
(\textbf{B}) $N_{\textrm {pop}}=3$, 
(\textbf{C}) $N_{\textrm {pop}}=4$, and 
(\textbf{D}) $N_{\textrm {pop}}=7$, for $N=10$ nodes per population in all cases.
We use black for one- and red for two-cluster states, and frequencies are shown relative to the first oscillator in the first one-cluster population.
For model description and other parameters, see \emph{Materials and Methods}.
 }
 }
 \label{fig:cn_np}
\end{figure}

\begin{figure}
\centering
\includegraphics[width= 0.7\columnwidth]{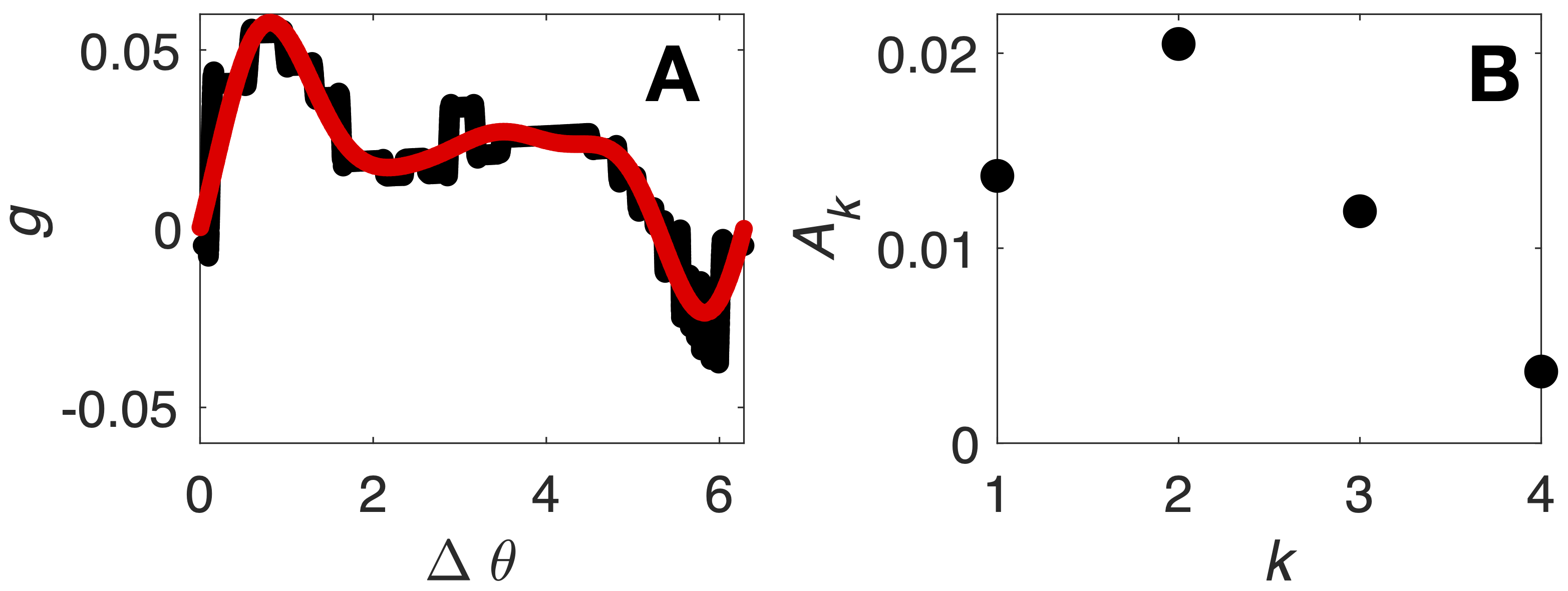}
\caption{\cb{\textbf{ Interaction function extracted numerically from the experiments. } 
The interaction function is estimated using data from the desynchronized state with the I-I configuration in Fig.\ \ref{fig:clusters}B.
We plot the instantaneous frequency of population  $\sigma$ (relative to the natural frequency) $\dot{\theta}_\sigma - \omega_\sigma$  versus $\Delta \theta$, where each population is represented by a single variable defined as the average over the currents in the population.
(\textbf{A})  Measured interaction function (black circles) and fitting to a Fourier expansion to fourth order (red curve).
(\textbf{B}) Magnitude of Fourier harmonics up to 4th order of the fitted interaction function in panel A.  
The resulting interaction function has a dominant second harmonic, with smaller magnitude first, third, and fourth harmonics. This function predicts  stable one-cluster ($\lambda_{\max} = -9.4\times10^{-2}$) and two-cluster 
($\lambda_{\max} = -1.0\times10^{-2}$) 
states with a frequency difference of $\Delta\Omega_{21}^\mathrm{u}=2.5\times10^{-2}$ rad/s. 
}
 }
 \label{fig:expg}
\end{figure}

\end{document}